\documentclass[aps,pra,twocolumn,amsfonts,amssymb,showkeys]{revtex4}
\usepackage{amsmath,mathrsfs} \usepackage{graphicx}
\usepackage{hyperref}

\def\Tr{\operatorname{Tr}} \def\d{\operatorname{d}}
\def\tr{\operatorname{Tr}} \def\d{\operatorname{d}}
\def\>{\rangle}\def\<{\langle} \def\sH{\mathcal{H}}
   \def\geq{\geqslant}\def\leq{\leqslant}

\def\N#1{\left|\!\left|{#1}\right|\!\right|} 
 
\def\supp{\mathsf{supp}}

 \def\states{\mathfrak{S}}
\def\id{\operatorname{id}}

\newcommand{\be}{\begin{equation}} \def\states{\mathfrak{S}}
  \newcommand{\ee}{\end{equation}} 
\newcommand{\eps}{{\varepsilon}} \newcommand{\bea}{\begin{eqnarray}}
  \newcommand{\F}{\mathsf{F}} \newcommand{\mD}{{\mathcal{D}}}
   \def\avg{\mathrm{avg}}
  \newcommand{\eea}{\end{eqnarray}}

\newtheorem{theorem}{Theorem}
\newtheorem{corollary}{Corollary}
\newtheorem{lemma}{Lemma}

\newtheorem{definition}{Definition}
\def\reff#1{(\ref{#1})}

\def\sS{\mathcal{S}}
\def\B{\mathfrak{b}}

\def\H{\widetilde{H}}
\def\I{\widetilde{I}^c}

\def\P{\mathfrak{p}}
\def\ent{\mathrm{ent}}

\begin{document}

\title{The quantum capacity of channels with arbitrarily
  correlated noise}

\author{Francesco Buscemi} \email{buscemi@statslab.cam.ac.uk}
\author{Nilanjana Datta} \email{n.datta@statslab.cam.ac.uk}
\affiliation{Statistical Laboratory, DPMMS, University of Cambridge,
  Cambridge CB3 0WB, UK}

\date{\today}

\begin{abstract}

  We study optimal rates for quantum communication over a single use
  of a channel, which itself can correspond to a finite number of uses
  of a channel with arbitrarily correlated noise. The corresponding
  capacity is often referred to as the \emph{one-shot} quantum
  capacity. In this paper, we prove bounds on the one-shot quantum
  capacity of an arbitrary channel. This allows us to compute the
  quantum capacity of a channel with arbitrarily correlated noise, in
  the limit of asymptotically many uses of the channel. In the memoryless case, we explicitly show that our results reduce to known expressions for the quantum capacity.

\end{abstract}


\maketitle

\section{Introduction}

In contrast to a classical channel which has a unique capacity, a
quantum channel has various distinct capacities. This is a consequence
of the greater flexibility in the use of a quantum channel.  As
regards transmission of information through it, the different
capacities arise from various factors: the nature of the transmitted
information (classical or quantum), the nature of the input states
(entangled or product states) the nature of the measurements done on
the outputs of the channel (collective or individual), the absence or
presence of any additional resource, e.g., prior shared entanglement
between sender and receiver, and whether they are allowed to
communicate classically with each other. The classical capacity of a
quantum channel under the constraint of product state inputs was shown
by Holevo \cite{holevo}, Schumacher and Westmoreland \cite{SW1} to be
given by the Holevo capacity of the channel. The capacity of a quantum
channel to transmit quantum information, in the absence of classical
communication and any additional resource, and without any constraint
on the inputs and the measurements, is called the quantum capacity of
the channel. It is known to be given by the regularized coherent
information \cite{lloyd, shor, devetak}.  A quantum channel can also
be used to generate entanglement between two parties, which can then
be used as a resource for teleportation. The corresponding capacity is
referred to as the entanglement generation capacity of the quantum
channel and is equivalent to the capacity of the channel for
transmitting quantum information~\cite{devetak}.

All these capacities were originally evaluated in the limit of asymptotically many uses of the channel, under the assumption that the noise acting on successive inputs to the channel is uncorrelated, i.e., under the assumption that the channel is \emph{memoryless}.  In reality, however, this assumption, and the consideration of an asymptotic scenario, is not necessarily justified. It is hence of importance to evaluate both $(i)$ bounds on the {\em{one-shot
    capacities}} of a quantum channel, that is its capacities for a
finite number uses or even a single use, as well as $(ii)$ the
capacity of an arbitrary sequence of channels, possibly with
memory. Both these issues are addressed in this paper.

For an arbitrary quantum channel, it is not in general possible to achieve perfect information transmission or entanglement generation over a single use or a finite number of uses. Hence, one needs to allow for a non-zero probability of error. This leads us to consider the capacities under the constraint that the probability of error is at most $\eps$, for a given $\eps\ge0$.

In this paper we consider the following protocol, which we call
{\em{entanglement transmission}}~\cite{shordevetak}. Let $\Phi$ be a quantum channel, let
${\cal{H}}_M$ be a subspace of its input Hilbert space, and let $\eps$
be a fixed positive constant. Suppose Alice prepares a maximally
entangled state $|\Psi^+\> \in {\cal{H}}_M \otimes {\cal{H}}_{M'}$,
where ${\cal{H}}_{M'} \simeq {\cal{H}}_{M}$, and sends the part $M$
through the channel $\Phi$ to Bob. Bob is allowed to do any decoding
operation (completely positive trace-preserving map) on the state that
he receives. The final objective is for Alice and Bob to end up with a
shared state which is nearly maximally entangled over ${\cal{H}}_M
\otimes {\cal{H}}_{M'}$, its overlap with $|\Psi^+\>$ being at least
$(1-\eps)$. In this protocol, there is no classical communication
allowed between Alice and Bob. For a given
$\eps\ge0$, let $Q_{\ent}(\Phi; \eps)$ denote the \emph{one-shot
  capacity of entanglement transmission}. In this paper we prove that this
capacity is expressible in terms of a generalization of relative R\'enyi
entropy of order 0. Our results also yield a characterization of the
\emph{one-shot quantum capacity} of the channel. This is because it
can be shown that the one-shot capacity of transmission of {\em{any}}
quantum state by the channel, evaluated under the condition that the
minimum fidelity of the channel is at most $(1 - \eps)$, for a given
$\eps\ge 0$, is bounded above by $Q_{\ent}(\Phi; \eps)$, and
bounded below by $Q_{\ent}(\Phi; \eps/2) -1$ (see Section~\ref{sec:5}).

By the Stinespring Dilation Theorem~\cite{stine}, the action of a quantum channel creates correlations between the sender, the receiver, and the environment interacting with the input. Faithful transmission of quantum information requires a decoupling of the state of the environment from that of the sender (see the special issue~\cite{osid}). In~\cite{dec-capacity}, a lower bound to the accuracy with which this decoupling can be achieved in a single use of the channel, was obtained. Here we go a step further and evaluate bounds on the one-shot capacity. In evaluating the lower bound, we employ an inequality, given by Lemma~\ref{lemma:krs}, relating the decoupling accuracy to the decoding fidelity. To obtain the upper bound we instead generalize the standard arguments relying on the quantum data-processing inequality~\cite{devetak,nielsen}. Moreover, in the limit of asymptotically many uses of a memoryless channel, we prove, without explicitly resorting to any typicality argument, that each of these bounds converge independently to the familiar expression of the quantum capacity given by the regularized coherent information~\cite{lloyd, shor, devetak}. For the important case of an arbitrary sequence of channels, possibly with memory, our one-shot result yields the asymptotic quantum capacity in the Information Spectrum framework~\cite{info-spect,hayashi-naga}.

We start the paper with some definitions and notations in Section~\ref{prelim}, including that of quasi-entropies, which play a pivotal role in our analysis. In Section~\ref{fidelities} we introduce the protocol of entanglement transmission, and define its fidelity and the corresponding one-shot capacity. Our main result is given by Theorem~\ref{thm_main} of Section~\ref{bounds}. In Section~\ref{sec:5} we relate the one-shot entanglement transmission capacity with the one-shot quantum capacity. The tools used for the proof of Theorem~\ref{thm_main} are given in Section~\ref{sec:prereq}, with the proof itself presented in Section~\ref{sec:proof}. Further, in Section~\ref{multiple}, we consider a sequence of arbitrary channels, with or without memory, and derive an expression for its asymptotic quantum capacity. When the channels in the sequence are memoryless, we recover known expressions for quantum capacity given in terms of the regularized coherent information. We conclude with a discussion of our results in Section~\ref{discussion}.


\section{Definitions and notations}
\label{prelim}

\subsection{Mathematical preliminaries}

Let ${\cal B}(\sH)$ denote the algebra of linear operators acting on a
finite--dimensional Hilbert space $\sH$ and let $\states(\sH)$ denote
the set of positive operators of unit trace (states) acting on
$\sH$. A quantum channel is given by a completely positive
trace--preserving (CPTP) map $\Phi: {\cal B}({\cal H}_A) \mapsto {\cal
  B}({\cal H}_B)$, where ${\cal H}_A$ and ${\cal H}_B$ are the input and output Hilbert spaces of the channel. Moreover, for any given subspace $\sS\subseteq\sH_A$, we define the restriction of the channel $\Phi$ to the subspace $\sS$ as $\Phi|_\sS(\rho):=\Phi(\Pi_\sS\rho\Pi_\sS)$, for any $\rho\in\mathcal{B}(\sH_A)$, with $\Pi_\sS$ being the projector onto $\sS$. Notice that $\Phi|_\sS$ is itself a CPTP-map $\Phi|_\sS:\mathcal{B}(\sS)\mapsto\mathcal{B}(\sH_B)$. Throughout this paper we restrict our considerations to finite-dimensional Hilbert spaces, and we take the logarithm to base $2$.

For given orthonormal bases $\{|i^A\rangle\}_{i=1}^d$ and
$\{|i^B\rangle\}_{i=1}^d$ in isomorphic Hilbert spaces
${\cal{H}}_A\simeq{\cal{H}}_B\simeq\sH$ of dimension $d$, we define a
maximally entangled state (MES) of rank $m \le d$ to be
\begin{equation}\label{MES-m}
|\Psi_m^{AB}\>= \frac{1}{\sqrt{m}} \sum_{i=1}^m |i^A\rangle\otimes |i^B\rangle.
\end{equation}
When $m=d$, for any given operator $A\in\mathcal{B(\sH)}$, the
following relation can be shown by direct inspection:
\begin{equation}\label{ricochet}
  (A\otimes\openone)|\Psi_d^{AB}\>=(\openone\otimes A^T)|\Psi_d^{AB}\>,
\end{equation}
where $\openone$ denotes the identity operator, and $A^T$ denotes the
transposition with respect to the basis fixed by
eq.~\reff{MES-m}. Moreover, for any given pure state $|\phi\>$, we
denote the projector $|\phi\>\<\phi|$ simply as $\phi$.

The trace distance between two operators $A$ and $B$ is given by
\begin{equation}\nonumber
  \N{A-B}_1 := \tr\bigl[\{A \ge B\}(A-B)\bigr] - \tr\bigl[\{A <
  B\}(A-B)\bigr],
\end{equation}
where $\{A\ge B\}$ denotes the projector on the
subspace where the operator $(A-B)$ is non-negative, and
$\{A<B\}:=\openone-\{A\ge B\}$. The fidelity of two states $\rho$ and
$\sigma$ is defined as
\begin{equation}\label{fidelity-aaa}
F(\rho, \sigma):= \tr \sqrt{\sqrt{\rho} \sigma \sqrt{\rho}}
=\N{\sqrt{\rho}\sqrt{\sigma}}_1.
\end{equation}
The trace distance between two states $\rho$ and $\sigma$ is related
to the fidelity $F(\rho, \sigma)$ as follows (see
e.~g.~\cite{nielsen}):
\begin{equation}
  1-F(\rho,\sigma) \leq \frac{1}{2} \N{\rho -
    \sigma}_1 \leq \sqrt{1-F^2(\rho, \sigma)},
\label{fidelity}
\end{equation}
where we use the notation $F^2(\rho, \sigma) = \bigl(F(\rho,\sigma)
\bigr)^2$. We also use the following results:

\begin{lemma}[\cite{bowen-datta}]\label{bowen}
  For any self-adjoint operators $A$ and $B$, and any positive
  operator $0\le P\le\openone$,
\begin{equation}\nonumber
\Tr[P(A-B)]\le\Tr[\{A\ge B\}(A-B)]
\end{equation}
and
\begin{equation}\nonumber
  \Tr[P(A-B)]\ge\Tr[\{A< B\}(A-B)].\ \square
\end{equation}
\end{lemma}

\begin{lemma}[Gentle measurement lemma~\cite{winter99,ogawanagaoka02}]
\label{gm}
For a state $\rho\in\states(\sH)$ and operator $0\le
\Lambda\le\openone$, if $\Tr(\rho\ \Lambda) \ge 1 - \delta$, then
$$\N{\rho -   {\sqrt{\Lambda}}\rho{\sqrt{\Lambda}}}_1 \le {2\sqrt{\delta}}.$$
The same holds if $\rho$ is a subnormalized density operator. $\square$
\end{lemma}

\begin{lemma}[\cite{thesis}]\label{lemma:tr-hs}
  For any self-adjoint operator $X$ and any positive operator $\xi>0$,
  we have
  \begin{equation}\label{orig}
\begin{split}
    \N{X}_1^2&\le\Tr[\xi]\Tr\left[X\xi^{-1/2}X\xi^{-1/2}\right]\\
&\le\Tr[\xi]\Tr\left[X^2\xi^{-1}\right].\ \square
\end{split}
\end{equation}
\end{lemma}

\noindent{\bf Proof.} The first inequality in~\reff{orig} was proved
in~\cite{thesis}. The second one simply follows as an application of
the Cauchy-Schwarz inequality, that is,
\begin{equation}\nonumber
\begin{split}
&\phantom{\ge}\Tr\left[X\xi^{-1/2}X\xi^{-1/2}\right]\\
&\le\sqrt{\Tr\left[\left(X\xi^{-1/2}\right)\left(X\xi^{-1/2}\right)^\dag\right]}\sqrt{\Tr\left[\left(X\xi^{-1/2}\right)^\dag\left(X\xi^{-1/2}\right)\right]}\\
&=\Tr\left[X^2\xi^{-1}\right].\ \blacksquare
\end{split}
\end{equation}

\begin{lemma}\label{lemma:krs}
Given a tripartite pure state $|\Omega^{RBE}\>\in\sH_R\otimes \sH_B\otimes \sH_E$, let $\omega^{RB}$, $\omega^{RE}$, $\omega^R$, and $\omega^E$ be its reduced states. Then
\begin{equation}
\begin{split}
&F^2(\omega^{RE},\omega^R\otimes\omega^E)\\
\le&\max_{\mathcal{D}}F^2((\id_R\otimes\mD_B)(\omega^{RB}),\Psi^{RA}),
\end{split}
\end{equation}
where $|\Psi^{RA}\>\in\sH_R\otimes\sH_A$ is some fixed purification of $\omega^R$ and $\mathcal{D}:\mathcal{B}(\sH_B)\mapsto\mathcal{B}(\sH_A)$ denotes a CPTP map. $\square$
\end{lemma}

\noindent{\bf Proof.} Fix some purification $|\chi^{EA'}\>\in\sH_E\otimes\sH_{A'}$ of $\omega^E$. Then, for the fixed purification $|\Psi^{RA}\>$ of $\omega^R$, we have, by Uhlmann's theorem~\cite{uhlmann}, the monotonicity of the fidelity under partial trace, and Stinespring's Dilation Theorem~\cite{stine},
\begin{equation}\label{eq:proof_4}
  \begin{split}
&F^2(\omega^{RE},\omega^R\otimes\omega^E)\\
=&\max_{|\varphi^{REAA'}\>\atop{\Tr_{AA'}[\varphi^{REAA'}]=\omega^{RE}}}F^2(\varphi^{REAA'},\Psi^{RA}\otimes\chi^{EA'})\\
=&\max_{V:B\to AA'\atop{V^\dag V=\openone_B}}F^2\left((\openone^{RE}\otimes V_B)\Omega^{RBE}(\openone^{RE}\otimes V_B^\dag),\Psi^{RA}\otimes\chi^{EA'}\right)\\
\le&\max_{\mD}F^2\left((\id_R\otimes\mD_B)(\omega^{RB}),\Psi^{RA}\right),
\end{split}
\end{equation}
where $\mD:\mathcal{B}(\sH_B)\mapsto\mathcal{B}(\sH_A)$ denotes a CPTP map. In the second equality of~(\ref{eq:proof_4}) we also used the well-known fact that all possible purifications of a given mixed state ($\omega^{RE}$, in our case) are related by some local isometry acting on the purifying system only (i.e. subsystem $B$). $\blacksquare$

\subsection{Quasi-entropies and coherent information}\label{entropies}

For any $\rho,\sigma\ge 0$ and any $0\le P\le\openone$, the {\em{quantum
    relative quasi-entropy of order $\alpha$}}~\cite{petz}, for $\alpha\in(0,\infty)\backslash\{1\}$, is defined as
\begin{equation}\label{quasi-ent}
S_\alpha^P(\rho\|\sigma):=\frac{1}{\alpha-1}\log\Tr[\sqrt{P}\rho^\alpha\sqrt{P}\sigma^{1-\alpha}].
\end{equation}
Notice that for $P=\openone$, the quasi-entropy defined above reduces to the well-known R\'enyi relative entropy of order $\alpha$.

In this paper, in particular, the quasi-entropy of order 0, namely,
\begin{equation}
S_0^P(\rho\|\sigma):=\lim_{\alpha\searrow 0}S_\alpha^P(\rho\|\sigma),
\end{equation}
plays an important role. Note that
\begin{equation}
S_0^P(\rho\|\sigma)=-\log\Tr[\sqrt{P}\Pi_\rho\sqrt{P}\ \sigma],
\end{equation}
where $\Pi_\rho$ denotes the projector onto the support of $\rho$. Our main result, Theorem~\ref{thm_main}, is expressible in terms of two ``smoothed'' quantities, which are derived from the quasi-entropy of order 0, for any $\delta\ge 0$, as
\begin{equation}\label{eq:i}
I^c_{0,\delta}(\rho^{AB}):=\max_{\bar\rho^{AB}\in \B(\rho^{AB};\delta)}\min_{\sigma^B\in\states(\sH_B)}S_0^{\openone}(\bar\rho^{AB}\|\openone_A\otimes\sigma^B),
\end{equation}
and
\begin{equation}\label{eq:itilda}
\I_{0,\delta}(\rho^{AB}):=\max_{P\in \P(\rho^{AB};\delta)}\min_{\sigma^B\in\states(\sH_B)}S_0^P(\rho^{AB}\|\openone_A\otimes\sigma^B),
\end{equation}
where
\begin{equation}
 \B(\rho;\delta):=\{\sigma:\sigma\ge 0,\ \Tr[\sigma]\le 1,\ F^2(\rho,\sigma)\ge1-\delta^2\},\label{ball}
\end{equation}
and
\begin{equation}
  \P(\rho;\delta):=\{P:0\le P\le\openone,\ \Tr[P\rho]\ge1-\delta\}.\label{P-ball}
\end{equation}
(Note that, in~\reff{ball}, the definition of fidelity~\reff{fidelity-aaa} has been naturally extended to subnormalized density operators.)
Such smoothed quantities are needed in order to allow for a finite accuracy (i.e. non-zero error) in the protocol, which is a natural requirement in the one-shot regime. Their properties are discussed in detail in Section~\ref{smooth}.

\section{The protocol: entanglement transmission}\label{fidelities}

As mentioned in the Introduction, we consider the protocol of {\em{entanglement transmission}}~\cite{shordevetak}: Given a quantum channel $\Phi:{\cal
  B}(\sH_A)\mapsto{\cal B}(\sH_B)$, let ${\cal{H}}_M$ be an $m$-dimensional subspace of its input Hilbert space, and let $\eps$ be a fixed positive constant. Alice prepares a maximally entangled state $|\Psi^{M'M}_m\>\in {\cal{H}}_{M'} \otimes {\cal{H}}_{M}$, where ${\cal{H}}_{M'} \simeq {\cal{H}}_{M}$, and sends the part $M$ through the channel $\Phi$ to Bob. Bob is allowed to do any decoding operation (CPTP map) on the state that he receives. The final objective is for Alice and Bob to end up with a shared state which is nearly maximally entangled over ${\cal{H}}_{M'} \otimes {\cal{H}}_{M}$, its overlap with $|\Psi^{M'M}_m\>$ being at least $(1-\eps)$. There is no classical communication possible between Alice and Bob. Within this scenario, for any positive integer $m$, the efficiency of the channel $\Phi$ in transmitting entanglement, is given in terms of the fidelity defined below:

\begin{definition}[Entanglement transmission fidelity] Let a channel $\Phi:{\cal
    B}(\sH_A)\mapsto{\cal B}(\sH_B)$ be given. For any given positive integer $m\le\dim\sH_A$, we define the entanglement transmission fidelity of $\Phi$ as
\begin{equation}
\begin{split}
  &\F_\ent(\Phi;m)\\
:=&\max_{\sH_M\subseteq\sH_A\atop{\dim\sH_M=m}}\max_{\mD}\ \<\Psi^{M'M}_m|(\id\otimes\mD\circ\Phi)(\Psi^{M'M}_m)|\Psi^{M'M}_m\>,
\label{fsub}
\end{split}
\end{equation}
where $\mD:{\cal B}(\sH_B)\mapsto{\cal B}(\sH_A)$ is a decoding
CPTP-map.\label{def:ent-fid}  $\square$
\end{definition}

We can now define an achievable rate as follows:
\begin{definition}[$\eps$-achievable rate]
  Given a channel $\Phi:{\cal B}(\sH_A)\mapsto{\cal B}(\sH_B)$ and a
  real number $\eps\ge 0$, any $R=\log m$, $m\in\mathbb{N}$,
  is an $\eps$-achievable rate, if
\begin{equation}\nonumber
\F_\ent(\Phi;m)\ge 1-\eps.\  \square
\end{equation}\label{eachrate}
\end{definition}

This leads to the definition of
the one-shot capacity of entanglement transmission:

\begin{definition}[One-shot capacity]
  Given a quantum channel $\Phi:{\cal B}(\sH_A)\mapsto{\cal B}(\sH_B)$
  and a real number $\eps\ge 0$, the one-shot capacity of entanglement transmission of $\Phi$
  is defined as
\begin{equation}\nonumber 
  Q_{\ent}(\Phi;\eps):=\max\{R: R\ \mathrm{is}\ \eps\mathrm{-achievable}\}.\  \square
\end{equation}
\end{definition}

\section{Main result: one-shot entanglement transmission capacity}\label{bounds}

Given a Hilbert space $\sH_A$ with $d:=\dim\sH_A$, let $\sH_R$ be
isomorphic to $\sH_A$, and fix a basis $\{|i^R\>\}_{i=1}^d$ for
$\sH_R$. Then, for any given subspace ${\cal S}\subseteq\sH_A$ of
dimension $s$, we construct the maximally entangled state of rank $s$
in $\sH_R\otimes\sH_A$ as
\begin{equation}\label{sub-MES}
  |\Psi_\sS^{RA}\>:=\frac 1{\sqrt{s}}\sum_{i=1}^s|i^R\>\otimes|\varsigma_i^A\>,
\end{equation}
where $\{|\varsigma_i^A\>\}_{i=1}^s$ is an orthonormal basis for
$\sS$. Now, given a channel
$\Phi:\mathcal{B}(\sH_A)\mapsto\mathcal{B}(\sH_B)$, let
$V_\Phi^A:\sH_A\mapsto\sH_B\otimes\sH_E$ be a Stinespring isometry
realizing the channel $\Phi$ as
\begin{equation}\nonumber
  \Phi(\rho)=\Tr_E[V_\Phi\rho V_\Phi^\dag],
\end{equation}
for any $\rho\in\states({\sH_A})$. For any subspace
$\sS\subseteq\sH_A$, from eq.~\reff{sub-MES}, we define the tripartite
pure state
\begin{equation}\label{o-rbe}
  |\Omega^{RBE}_\sS\>:=(\openone_R\otimes V^A_\Phi)|\Psi^{RA}_\sS\>.
\end{equation}
We then define $\omega^{RB}_\sS:=\Tr_E[\Omega^{RBE}_\sS]$ and
$\omega^{RE}_\sS:=\Tr_B[\Omega^{RBE}_\sS]$ to be its reduced
states. Our main result is stated in Theorem \ref{thm_main} below.

\begin{theorem}
\label{thm_main} 
For any $\eps\ge 0$, the one-shot capacity of entanglement transmission
for a quantum channel $\Phi:\mathcal{B}(\sH_A)\mapsto \mathcal{B}(\sH_B)$,
$Q_\ent(\Phi;\eps)$, satisfies the following bounds:
\begin{equation}
\begin{split}
  \phantom{\frac 1n}&\max_{\sS\subseteq\sH_A}I^c_{0,\eps/8}(\omega^{RB}_\sS)+\log\left[\frac
    1d+\frac{\eps^2}{4}\right]-\Delta\\
  \phantom{\frac 1n}\le&\ Q_\ent(\Phi;\eps)\\
  \phantom{\frac 1n}\le&\max_{\sS\subseteq\sH_A}\I_{0,2\sqrt{\eps}}(\omega^{RB}_\sS),
\label{eqcorr3}
\end{split}
\end{equation}
where $d:=\dim\sH_A$, $I^c_{0,\eps/8}(\omega^{RB}_\sS)$
and $\I_{0,2\sqrt{\eps}}(\omega^{RB}_\sS)$ are
the smoothed $0$-coherent informations
defined, respectively, by~\reff{eq:i} and~\reff{eq:itilda}, and $0\le\Delta\le 1$ is included to ensure that the lower bound is equal to the logarithm of a positive integer. $\square$
\end{theorem}

\noindent{\bf Remark.} Given a positive real $x$, for $x-\Delta$ to be the logarithm of a positive integer, we must have $\Delta\equiv\Delta(x):=x-\log\left\lfloor2^x\right\rfloor$, where $\lfloor y\rfloor$ denotes the largest integer less than or equal to $y$. It can be shown that $0\le\Delta(x)\le 1$ for all $x\ge 0$, and that $\Delta(x)$ decreases rapidly as $x$ increases.

\section{One-shot quantum capacity}\label{sec:5}

It is interesting to compare the entanglement transmission fidelity of a quantum channel with the minimum output fidelity defined below:

\begin{definition}[Minimum output fidelity]\label{def:min-fid}
  Let a channel $\Phi:{\cal B}(\sH_A)\mapsto{\cal B}(\sH_B)$ be given. For any given positive integer $m$, we define the minimum output fidelity of $\Phi$ as
\begin{equation}\nonumber
\F_\mathrm{min}(\Phi;m):=\max_{\sH_M\subseteq\sH_A\atop{\dim\sH_M=m}}\max_{\mD}\min_{|\phi\>\in\sH_M}\ \<\phi|(\mD\circ\Phi)(\phi)|\phi\>,
\end{equation}
where $\mD:{\cal B}(\sH_B)\mapsto{\cal B}(\sH_A)$ is a decoding CPTP-map. $\square$
\end{definition}

\noindent{\bf Remark.} Note that Definitions~\ref{def:ent-fid} and~\ref{def:min-fid} include an optimization over all decoding operations. Hence they provide a measure of how well the effect of the noise in the channel can be corrected. This is in contrast to the definitions of fidelities used in~\cite{bkn,dennis} which provide a measure of the ``distance'' of a given channel from the trivial (identity) channel.\bigskip

The minimum output fidelity is related to the entanglement transmission fidelity through the following lemma~\cite{bkn,dennis}:
\begin{lemma}[Pruning Lemma]\label{lemma:purging}
  Let a channel $\Phi:{\cal B}(\sH_A)\mapsto{\cal B}(\sH_B)$ be given.
  Then , for any positive integer $m$,
\begin{equation}\nonumber
\F_\mathrm{min}(\Phi;m/2)\ge 1- 2\left[1-\F_\ent(\Phi;m)\right].\ \square
\end{equation}
\end{lemma}

Analogously to what we did for the entanglement transmission fidelity, one could also define the one-shot capacity with respect to the fidelity $\F_{\min}$ as follows:
\begin{equation}
Q_{\min}(\Phi;\eps):=\max\{\log m:F_{\min}(\Phi;m)\ge 1-\eps\}.
\end{equation}

\noindent{\bf Remark.} Note that quantum capacity is traditionally
defined with respect to the minimum output fidelity
$\F_{\min}$~\cite{devetak}. Hence, we define $Q_{\min}(\Phi;\eps)$ to
be the one-shot quantum capacity of a channel $\Phi$, for any
$\eps\ge0$.\bigskip

The following corollary, derived from Lemma~\ref{lemma:purging}, allows us to relate the one-shot entanglement transmission capacity $Q_{\ent}(\Phi;\eps)$ to the one-shot quantum capacity:
\begin{corollary} \label{corr1}
Given a quantum channel $\Phi:{\cal B}(\sH_A)\mapsto{\cal B}(\sH_B)$ and a real number $\eps>0$, 
\begin{equation}\nonumber
Q_{\ent}(\Phi;\eps)-1\le Q_{\min}(\Phi;2\eps)\le Q_{\ent}(\Phi;4 \eps).\ \square
\end{equation}
\end{corollary}
{\bf{Proof.}}
The lower bound follows directly from the Pruning Lemma. To prove the upper bound we resort 
to another frequently used fidelity, namely, the {\em{average fidelity}}:
\begin{equation}\nonumber
  {\F}_{\avg}(\Phi;m):=\max_{\sH_M\subseteq\sH_A\atop{\dim\sH_M=m}} \max_{\mD}\ \int\d\phi\ \<\phi|(\mD\circ\Phi)(\phi)|\phi\>,
\end{equation}
where $\d\phi$ is the normalized unitarily invariant measure over pure
states in $\sH_M$, and $\mD:{\cal B}(\sH_B)\mapsto{\cal B}(\sH_A)$ is a
decoding CPTP-map.

In \cite{sub-vs-avg} the relation of the above fidelity to the entanglement transmission fidelity
was shown to be given by:
\begin{equation}\nonumber
\F_{\avg}(\Phi;m)=\frac{m\cdot\F_{\ent}(\Phi;m)+1}{m+1},
\end{equation}
while clearly, by definition, $ \F_{\min}(\Phi;m)\le  \F_{\avg}(\Phi;m)$.
Hence, if $\F_{\min}(\Phi;m)\ge 1-\eps'$, then 
\begin{eqnarray}
\F_{\ent}(\Phi;m) & \ge & \frac{(m+1)(1-\eps') - 1}{m}\nonumber\\
&=& 1 - \frac{m+1}{m}\eps' \ge 1 - 2 \eps'.\ \blacksquare\nonumber
\end{eqnarray}

Note that, due to Corollary~\ref{corr1}, Theorem~\ref{thm_main} provides bounds on the one-shot quantum capacity of a channel as well.

\section{Tools used in the proof}\label{sec:prereq}

The proof of Theorem~\ref{thm_main} relies on the properties of various entropic quantities derived from the relative quasi-entropies defined in Section~\ref{entropies}.

\subsection{Quantum entropies}

Let us first consider the relative R\'enyi entropy of order $\alpha$, which as mentioned before, is obtained from the quasi-entropy~(\ref{quasi-ent}) by setting $P=\openone$. (In the following, when $P=\openone$, we will drop the exponent in writing relative R\'enyi entropies, for sake of notational simplicity.) It is known that
\begin{equation}\nonumber
S_1(\rho\|\sigma):=\lim_{\alpha\nearrow 1}S_\alpha(\rho\|\sigma)=S(\rho\|\sigma),
\end{equation}
where $S(\rho\|\sigma)$ is the usual quantum relative entropy defined as
\begin{equation}\label{q-rel}
S(\rho\|\sigma):=\left\{
\begin{split}
&\Tr[\rho\log\rho-\rho\log\sigma],\textrm{ if }\supp\ \rho\subseteq\supp\ \sigma\\
&+\infty,\textrm{ otherwise}.
\end{split}
\right.
\end{equation}
From this, one derives the von Neumann entropy $S(\rho)$ of a state
$\rho$ as $S(\rho)=-S(\rho\|\openone)$.
We make use of the following
lemma in the sequel:
\begin{lemma}\label{lemma:qmi_as_min}
  Given a state $\rho^{AB} \in {\cal{H}}_A \otimes {\cal{H}}_B$, let
  $\rho^{A}:=\Tr_{B}[\rho^{AB}]$ and
  $\rho^{B}:=\Tr_{A}[\rho^{AB}]$. Then, for any operator $\sigma^A\ge
  0$ with $\supp\sigma^A\supseteq\supp\rho^A$,
\begin{equation}\nonumber
  \min_{\xi^B\ge 0}S(\rho^{AB}\|\sigma^A\otimes\xi^B)=S(\rho^{AB}\|\sigma^A\otimes\rho^B).
\end{equation}
This implies, in particular, that, for any state $\rho^{AB}$,
\begin{equation}\nonumber
  \min_{\xi^B\ge 0}S(\rho^{AB}\|\openone_A\otimes\xi^B)=S(\rho^{AB}\|\openone_A\otimes\rho^B),
\end{equation}
and
\begin{equation}\label{eq:qmi_as_min}
  \min_{\omega^A,\xi^B\ge 0}S(\rho^{AB}\|\omega^A\otimes\xi^B)=S(\rho^{AB}\|\rho^A\otimes\rho^B).\ \square
\end{equation}
\end{lemma} {\bf Proof.} Here we only prove eq.~(\ref{eq:qmi_as_min}).
The rest of the lemma can be proved exactly along the same lines. By
definition, we have that
\begin{equation}\nonumber
S(\rho^{AB}\|\omega^A\otimes\xi^B)=\Tr[\rho^{AB}\log\rho^{AB}]-\Tr[\rho^{AB}\log(\omega^A\otimes\xi^B)].
\end{equation}
Since $\log(\omega^A\otimes\xi^B)= (\log\omega^A)\otimes\openone_B+
\openone_A\otimes(\log\xi^B)$, we can rewrite
\begin{equation}\nonumber
\begin{split}
  &\phantom{=}S(\rho^{AB}\|\omega^A\otimes\xi^B)\\
&=\Tr[\rho^{AB}\log\rho^{AB}]-\Tr[\rho^A\log\omega^A]-\Tr[\rho^B\log\xi^B].
\end{split}
\end{equation}
Now, since for all $\rho$ and $\sigma$,
\begin{equation}\nonumber
  0\le S(\rho\|\sigma)=\Tr[\rho\log\rho]-\Tr[\rho\log\sigma],
\end{equation}
we have that
\begin{equation}\nonumber
\Tr[\rho\log\rho]\ge\Tr[\rho\log\sigma],
\end{equation}
which implies that
\begin{equation}\nonumber
\begin{split}
&\phantom{\ge}  S(\rho^{AB}\|\omega^A\otimes\xi^B)\\
&\ge\Tr[\rho^{AB}\log\rho^{AB}]-\Tr[\rho^A\log\rho^A]-\Tr[\rho^B\log\rho^B]\\
  &=S(\rho^{AB}\|\rho^A\otimes\rho^B).\ \blacksquare
\end{split}
\end{equation}\bigskip

Recently, a generalized relative entropy, namely the max-relative entropy $D_{\max}$, was introduced
in~\cite{nila}. For a state $\rho$ and an operator $\sigma\ge 0$,
\begin{equation}\nonumber
\begin{split}
  D_{\max}(\rho\|\sigma):&=\log\min\{\lambda:\rho\le\lambda\sigma\}\\
  &=\log\lambda_{\max}(\sigma^{-1/2}\rho\sigma^{-1/2}),
\end{split}
\end{equation}
$\lambda_{\max}(X)$ denoting the maximum eigenvalue of the operator
$X$. Even though for commuting $\rho$ and $\sigma$,
$D_{\max}(\rho\|\sigma)=\lim_{\alpha\to\infty}S_\alpha(\rho\|\sigma)$,
this identity does not hold in general~\cite{milan}. We can however
easily prove the following property:
\begin{lemma}\label{lemma:s2dmax}
  For any $\rho,\sigma\ge 0$ with $\Tr[\rho]\le 1$, we have
\begin{equation}\nonumber
  S_2(\rho\|\sigma)\le D_{\max}(\rho\|\sigma).\ \square
\end{equation}
\end{lemma} {\bf Proof.} By definition,
$2^{S_2(\rho\|\sigma)}=\Tr[\rho^2\sigma^{-1}]$. By noticing that, for
any Hermitian operator $X$ and any subnormalized state $\rho$,
$\Tr[\rho X]\le\lambda_{\max}(X)$, we obtain that
$\Tr[\rho^2\sigma^{-1}]= \Tr[\rho(\rho^{1/2}\sigma^{-1}\rho^{1/2})]\le
\lambda_{\max}(\rho^{1/2}\sigma^{-1}\rho^{1/2}) =
\lambda_{\max}(\sigma^{-1/2}\rho\sigma^{-1/2}) =
2^{D_{\max}(\rho\|\sigma)}$, where, in the last passage, we used the
fact that $\lambda_{\max}(A^\dag
A)=\lambda_{\max}(AA^\dag)$. $\blacksquare$\bigskip

Given an $\alpha-$relative R\'enyi entropy $S_\alpha(\rho\|\sigma)$,
for a bipartite $\rho=\rho^{AB}$, we define the corresponding
$\alpha$-conditional entropy as
\begin{equation}\label{eq:cond}
  H_\alpha(\rho^{AB}|\sigma^B):=-S_\alpha(\rho^{AB}\|\openone_A\otimes\sigma^B),
\end{equation}
and
\begin{equation}
\begin{split}
H_\alpha(\rho^{AB}|B):&=\max_{\sigma^B\in \states({\cal{H}}_B)}H_\alpha(\rho^{AB}|\sigma^B)\\
&=-\min_{\sigma^B\in\states({\cal{H}}_B)}S_\alpha(\rho^{AB}\|\openone_A\otimes\sigma^B).
\label{17}
\end{split}
\end{equation}
For a bipartite state $\rho^{AB}\in\states(\sH_A\otimes\sH_B)$, the
min-conditional entropy of $\rho^{AB}$ given $\sH_B$, denoted by
$H_{\min}(\rho^{AB}|B)$ and introduced by Renner~\cite{thesis}, is
relevant for the proof of our main result. It is obtainable from the
max-relative entropy as follows:
\begin{equation}\nonumber
  H_{\min}(\rho^{AB}|B):=-\min_{\sigma^B\in\states(\sH_B)}D_{\max}(\rho^{AB}\|\openone_A\otimes\sigma^B).
\end{equation}
Further, from the quantum relative
entropy~\reff{q-rel}, we define the quantum conditional entropy as
\begin{equation}\nonumber
H(\rho^{AB}|B)=-\min_{\sigma^B\in\states(\sH_B)}S(\rho^{AB}\|\openone_A\otimes\sigma^B),
\end{equation}
which, by Lemma~\ref{lemma:qmi_as_min}, satisfies
$H(\rho^{AB}|B)=H(\rho^{AB}|\rho^B)=S(\rho^{AB})-S(\rho^B)$. Finally,
given a bipartite state $\rho^{AB}$, its coherent information
$I^c (\rho^{AB})$ is defined as
\begin{equation}\label{coh}
  I^c (\rho^{AB}):=-H(\rho^{AB}|B)=S(\rho^B)-S(\rho^{AB}),
\end{equation}
and, by analogy,
\begin{equation}\nonumber
  I^c_\alpha (\rho^{AB}):=-H_\alpha(\rho^{AB}|B),
\end{equation}
for any $\alpha\in[0,\infty)$. Clearly, $I^c_1(\rho^{AB})=I^c (\rho^{AB})$.

\subsection{Smoothed entropies}\label{smooth}

As first noticed by Renner~\cite{thesis}, in order to allow for a finite accuracy in one-shot protocols,  it is necessary to
introduce smoothed entropies. We consider two different classes of smoothed entropies, namely the \emph{state-smoothed} and the \emph{operator-smoothed} entropies. The former was introduced by Renner~\cite{thesis}, while the latter arises naturally from the consideration of quasi-entropies.

\subsubsection{State-smoothed quantum entropies}

For any bipartite state $ \rho^{AB}\in
\states({\cal{H}}_A \otimes{\cal{H}}_B)$, smoothed conditional
entropies $H^\delta_{\min}(\rho^{AB}|B)$ and $H^\delta_0(\rho^{AB}|B)$
are defined for any $\delta\ge 0$ as
\begin{equation}\nonumber
\begin{split}
  &H^\delta_{\min}(\rho^{AB}|B):=\max_{\bar\rho^{AB}\in \B(\rho^{AB};\delta)}H_{\min}(\bar\rho^{AB}|B),\\
 &H^\delta_0(\rho^{AB}|B):=\min_{\bar\rho^{AB}\in \B(\rho^{AB};\delta)}H_0(\bar\rho^{AB}|B),
\end{split}
\end{equation}
where $\B(\rho^{AB};\delta)$ is the set defined in
eq.~\reff{ball}. For a bipartite $\rho^{AB}$, the smoothed
$\alpha$-conditional entropies $H^\delta_\alpha(\rho^{AB}|B)$ are then
defined, using~\reff{eq:cond} and~\reff{17}, as follows:
\begin{equation}\label{cond-smooth}
H_{\alpha}^\delta(\rho^{AB}|B):=\left\{
\begin{split}
  &\min_{\bar\rho^{AB}\in \B(\rho^{AB};\delta)}H_\alpha(\bar\rho^{AB}|B),\textrm{ for }0\le\alpha< 1\\
  &\max_{\bar\rho^{AB}\in \B(\rho^{AB};\delta)}H_\alpha(\bar\rho^{AB}|B),\textrm{ for }1<\alpha,\\
\end{split}
\right.
\end{equation}
and the corresponding smoothed $\alpha$-coherent information is
defined as
\begin{equation}\label{coh-smooth}
  I^c_{\alpha,\delta}(\rho^{AB}):=-H_\alpha^\delta(\rho^{AB}|B).
\end{equation}
For $\alpha=0$, this is identical to the definition~(\ref{eq:i}).

\subsubsection{Operator-smoothed quasi-entropies}

Given $\rho,\sigma\ge 0$ and an operator $0\le P\le\openone$, let us
consider the quantity
\begin{equation}\nonumber
  \psi_\alpha^P(\rho\|\sigma):=\log\Tr[\sqrt{P}\rho^\alpha\sqrt{P}\sigma^{1-\alpha}],\qquad\alpha>0.
\end{equation}
Note that $\psi_\alpha^P(\rho\|\sigma)$ is well-defined as long as
$\sigma^{1-\alpha}$ and $\sqrt{P}\rho^\alpha\sqrt{P}$ do not have
orthogonal supports. In the following, we shall assume this to be
true.

\begin{lemma}
  For any $\rho,\sigma\ge 0$, and any $0\le P\le\openone$, the
  function
\begin{equation}\nonumber
\alpha\mapsto\psi_\alpha^P(\rho\|\sigma)
\end{equation}
is convex for $\alpha>0$.  $\square$
\end{lemma}

\noindent{\bf Proof.} Let $\rho=\sum_ka_k|\gamma_k\>\<\gamma_k|$ and
$\sigma=\sum_lb_l|\beta_l\>\<\beta_l|$. Then,
\begin{equation}\nonumber
  \psi_\alpha^P(\rho\|\sigma)=\log\sum_{k,l}|c_{kl}|^2b_l\left(\frac{a_k}{b_l}\right)^\alpha,
\end{equation}
where $|c_{kl}|^2:=|\<\gamma_k|\sqrt{P}|\beta_l\>|^2$. By direct
inspection then,
\begin{equation}\nonumber
\frac{\d}{\d\alpha}\psi_\alpha^P(\rho\|\sigma)=\sum_{k,l}p_{kl}(\log a_k-\log b_l),
\end{equation}
where $p_{kl}$ is the probability distribution defined as
\begin{equation}\nonumber
  p_{kl}:=\frac{|c_{kl}|^2a_k^\alpha b_l^{1-\alpha}}{\sum_{k',l'}|c_{k'l'}|^2a_{k'}^\alpha b_{l'}^{1-\alpha}},
\end{equation}
and
\begin{equation}\nonumber
\begin{split}
 &\phantom{=} \frac{\d^2}{\d\alpha^2}\psi_\alpha^P(\rho\|\sigma)\\
&=\sum_{k,l}p_{kl}(\log a_k-\log b_l)^2-\left(\sum_{k,l}p_{kl}(\log a_k-\log b_l)\right)^2\\
&\ge0.
\end{split}
\end{equation}
Due to the positivity of its second derivative hence, the function
$\alpha\mapsto\psi_\alpha^P(\rho\|\sigma)$ is
convex. $\blacksquare$\bigskip

Note that the quantum relative quasi-entropy of order $\alpha$, $S_\alpha^P(\rho\|\sigma)$, can be equivalently written as
\begin{equation}\label{P-s}
  S^P_\alpha(\rho\|\sigma)=\frac{\psi_\alpha^P(\rho\|\sigma)}{\alpha-1}.
\end{equation}
It satisfies the following property:

\begin{lemma}\label{lemma:a-mono}
  For any $\rho,\sigma\ge 0$, and any $0\le P\le\openone$,
  $S^P_\alpha(\rho\|\sigma)$ is monotonically increasing in $\alpha$.  $\square$
\end{lemma}

\noindent{\bf Proof.} Due to convexity of
$\psi_\alpha^P(\rho\|\sigma)$, the function
\begin{equation}\nonumber
  \alpha\mapsto\frac{\psi_\alpha^P(\rho\|\sigma)-\psi_1^P(\rho\|\sigma)}{\alpha-1}
\end{equation}
is monotonically increasing in $\alpha$. Let us write, for our
convenience,
$f(\alpha):=\psi_\alpha^P(\rho\|\sigma)-\psi_1^P(\rho\|\sigma)$, and,
since
$\psi_1^P(\rho\|\sigma)=\log\Tr[\sqrt{P}\rho\sqrt{P}\ \Pi_\sigma]\le 0$,
let us put $-c:=\psi_1^P(\rho\|\sigma)\le 0$. Then, from monotonicity
of ${\displaystyle{\frac{f(x)+c}{x-1}}}$, we know that
\begin{equation}\nonumber
\begin{split}
  0&\le\frac{f'(x)(x-1)-(f(x)+c)}{(x-1)^2}\\
  &\le\frac{f'(x)(x-1)-f(x)}{(x-1)^2}.
\end{split}
\end{equation}
Since the second line is nothing but the derivative of
definition~\reff{P-s}, we proved the monotonicity of
$S_\alpha^P(\rho\|\sigma)$. $\blacksquare$

Let us now compute $S_1^P(\rho\|\sigma):=\lim_{\alpha\to
  1}S_\alpha^P(\rho\|\sigma)$: by l'H\^opital's rule,
\begin{equation}\label{S1P}
\begin{split}
&\phantom{=}\lim_{\alpha\to
    1}S_\alpha^P(\rho\|\sigma)=\left.\frac{\d}{\d\alpha}\psi_\alpha^P(\rho\|\sigma)\right|_{\alpha=1}\\
&=\frac{\Tr\left[\sqrt{P}\rho\log\rho\sqrt{P}\Pi_\sigma-\sqrt{P}\rho\sqrt{P}\log\sigma\right]}{\Tr[\sqrt{P}\rho\sqrt{P}\ \Pi_\sigma]}.
\end{split}
\end{equation}
This leads to the definition of the corresponding smoothed coherent
information:
\begin{equation}\label{eq:itildaone}
  \I_{1,\delta}(\rho^{AB}):=-\H_1^\delta(\rho^{AB}|B),
\end{equation}
where
\begin{equation}\nonumber
  \H_1^\delta(\rho^{AB}|B):=\min_{P\in\P(\rho^{AB};\delta)}\max_{\sigma^B\in\states(\sH_B)}[-S_1^P(\rho^{AB}\|\openone_A\otimes\sigma^B)].
\end{equation}
Analogously, for any bipartite state $\rho^{AB}$ and any $\delta\ge0$, the quantity $\I_{0,\delta}(\rho^{AB})$, given by~(\ref{eq:itilda}), is referred to as the operator-smoothed 0-coherent information. It is equivalently expressed as
\begin{equation}\label{new-I0}
\begin{split}
  -&\I_{0,\delta}(\rho^{AB})\\
=&\H_0^\delta(\rho^{AB}|B)\\
:=&\min_{P\in\P(\rho^{AB};\delta)}\max_{\sigma^B\in\states(\sH_B)}\log\Tr\left[\sqrt{P}\Pi_{\rho^{AB}}\sqrt{P}\
    (\openone_A\otimes\sigma^B)\right].
\end{split}
\end{equation}
The relation between $\I_{0,\delta}(\rho^{AB})$ defined in~(\ref{eq:itilda}) and
$\I_{1,\delta}(\rho^{AB})$ defined in~\reff{eq:itildaone} is provided by the following lemma:
\begin{lemma}\label{lemma:h-mono}
  For any $\rho^{AB}\in\states(\sH_A\otimes\sH_B)$ and any $\delta\ge
  0$,
\begin{equation}\nonumber
  \I_{0,\delta}(\rho^{AB})\le\I_{1,\delta}(\rho^{AB}).\ \square
\end{equation}
\end{lemma}

\noindent{\bf Proof.} Let $\bar P\in\P(\rho^{AB};\delta)$ be the
operator achieving $\I_{0,\delta}(\rho^{AB})$, and
let $\bar\sigma^B$ be the state achieving $\min_{\sigma^B}S_1^{\bar
  P}(\rho^{AB}\|\openone_A\otimes\sigma^B)$. Then,
\begin{equation}
\begin{split}
  \I_{1,\delta}(\rho^{AB})&\ge S_1^{\bar P}(\rho^{AB}\|\openone_A\otimes\bar\sigma^B)\\
  &\ge S_0^{\bar P}(\rho^{AB}\|\openone_A\otimes\bar\sigma^B)\\
  &\ge\min_{\sigma^B\in\states(\sH_B)}S_0^{\bar
      P}(\rho^{AB}\|\openone_A\otimes\sigma^B)\\
  &=\I_{0,\delta}(\rho^{AB}),
\end{split}
\end{equation}
where in the second line we used
Lemma~\ref{lemma:a-mono}. $\blacksquare$

\section{Proof of Theorem~\ref{thm_main}}\label{sec:proof}

\subsection{Proof of the lower bound in Theorem \ref{thm_main}}\label{proof}

The lower bound on the one-shot entanglement transmission capacity $Q_\ent(\Phi;\eps)$, for any fixed value $\eps\ge0$ of accuracy, is obtained by exploiting a lower bound on the entanglement transmission fidelity, which is derived below by the random coding method.

\subsubsection{Lower bound on entanglement transmission fidelity}\label{lbound}

The lower bound on the entanglement transmission fidelity is given by the following lemma:

\begin{lemma}
\label{thm_one}
Given a channel $\Phi:{\cal B}(\sH_A)\mapsto{\cal B}(\sH_B)$ and an
$s$-dimensional subspace $\sS\subseteq\sH_A$, consider the channel $\Phi|_\sS:\mathcal{B}(\sS)\mapsto\mathcal{B}(\sH_B)$ obtained by restricting $\Phi$ onto $\sS$, i.e. $\Phi|_\sS(\rho):=\Phi(\Pi_\sS\rho\Pi_\sS)$ for any $\rho\in\mathcal{B}(\sH_A)$, where $\Pi_\sS$ denotes the projector onto $\sS$. Then, for any $\delta\ge 0$ and any positive integer $m\le s$,
  \begin{equation}\label{eq:111}
    \F_{\ent}(\Phi|_\sS;m)\ge
    1-4\delta-\sqrt{m\left\{2^{I^c_{2,\delta}(\omega^{RE}_\sS)}-\frac
        1s\right\}},
  \end{equation}
  where $I^c_{2,\delta}(\omega^{RE}_\sS)$ is given
  by~\reff{coh-smooth} for $\alpha=2$.  $\square$

\end{lemma}

\noindent{\bf Remark.} From the theory of quantum error
correction~\cite{nielsen}, it is known that, for a channel noiseless
on $\sS$, $\omega^{RE}_\sS$, defined by~\reff{o-rbe} is a factorized state. Moreover, in our case,
$\omega^R_\sS:=\Tr_E[\omega^{RE}_\sS]=s^{-1}\sum_{i=1}^s|i\>\<i|^R$. As
shown in~\cite{KRS} by direct inspection, these two conditions imply
that $H_{\min}(\omega^{RE}_\sS|E)=\log
s$. On the other hand, from definition~\reff{17}, it follows that $H_0(\omega^{RE}_\sS|E)=\log s$. These two calculations, together with
the fact that $H_{\min}(\omega^{RE}_\sS|E)\le H_2(\omega^{RE}_\sS|E)\le H_0(\omega^{RE}_\sS|E)$, see~\cite{nila}, lead us to conclude
that also $H_2(\omega^{RE}_\sS|E)=\log s$, i.e., $I^c_2(\omega^{RE}_\sS)=-\log s$. Therefore, for any channel acting
noiselessly in $\sS$, $\F_{\ent}(\Phi|_\sS;m)=1$ for all
$m\le s$, as expected.\bigskip

\noindent{\bf Proof of Lemma~\ref{thm_one}.}
Fix the value of the positive integer $m\le s$. Then, starting from the pure state $|\Omega_\sS^{RBE}\>$ given by~\reff{o-rbe}, let us define
\begin{equation}\nonumber
  |\Omega_{m,g}^{RBE}\>:=\sqrt{\frac sm}(P_m^RU_g^R\otimes\openone_B\otimes\openone_E)|\Omega^{RBE}_\sS\>,
\end{equation}
where $U_g^R$ is a unitary representation of the element $g$ of the
group $\mathbb{SU}(s)$, and let
\begin{equation}\nonumber
P^R_m=\sum_{i=1}^m|i^R\>\<i^R|,
\end{equation}
the vectors $|i^R\>$, $i=1,\ldots,s$, being the same as in
eq.~(\ref{sub-MES}).  The reduced state $\Tr_B[\Omega^{RBE}_{m,g}]$
will be denoted as $\omega^{RE}_{m,g}$ (and analogously the
others). Notice that, by construction,
\begin{equation}\nonumber
  \omega^R_{m,g}=\tau^R_m:=\frac{P_m^R}{m}.
\end{equation}

The lower bound~\reff{eq:111} would follow if there exists a subspace $\sH_M\subseteq\sS$ of dimension $m$ which is transmitted with fidelity greater or equal to the right hand side of~\reff{eq:111}. One way to prove the existence of such a subspace is to show
that the \emph{group-averaged} fidelity,
$\overline{\F}(\sS,m)$ (defined below), is larger than that value:
\begin{equation}\label{avg}
\overline{\F}(\sS,m):=\int\d g\ \max_{\mD}F^2\left((\id_R\otimes\mD_B)(\omega^{RB}_{m,g}),\Psi^{RA}_{m,g}\right),
\end{equation}
where $|\Psi^{RA}_{m,g}\>:=\sqrt{\frac
  sm}(P_m^RU_g^R\otimes\openone_A)|\Psi^{RA}_\sS\>$, which is a MES of
rank $m$ due to~\reff{ricochet}. It is hence sufficient to compute a lower bound to
$\overline{\F}(\sS,m)$.

Using Lemma~\ref{lemma:krs}, we
have
\begin{equation}\nonumber
\overline{\F}(\sS,m) \ge \int\d g\
F^2\left(\omega^{RE}_{m,g},\tau^R_m\otimes\omega^E_{m,g}\right).
\end{equation}
Further, using the formula $F^2(\rho,\sigma)\ge
1-\N{\rho-\sigma}_1$, we have that
\begin{equation}\nonumber
  \overline{\F}(\sS,m)\ge 1-\int\d g\ \N{\omega^{RE}_{m,g}-\tau^R_m\otimes\omega^E_{m,g}}_1.
\end{equation}

Now, for any fixed $\delta\ge0$, let
$\bar\omega^{RE}\in\B(\omega^{RE}_\sS;\delta)$. Let us, moreover,
define $\bar\omega^{RE}_{m,g}:=\frac sm(P_M^RU_g^R\otimes\openone_E)
\bar\omega^{RE} (P_M^RU_g^R\otimes\openone_E)^\dag$. By the triangle
inequality, we have that
\begin{eqnarray}
  \phantom{\le}&\N{\omega^{RE}_{m,g}-\tau^R_m\otimes\omega^E_{m,g}}_1&\nonumber\\
  \le&\N{\bar\omega^{RE}_{m,g}-\tau^R_m\otimes\bar\omega_{m,g}^E}_1&+\N{\omega^{RE}_{m,g}-\bar\omega^{RE}_{m,g}}_1\nonumber\\
  &&+\N{\tau^R_m\otimes\bar\omega_{m,g}^E-\tau^R_m\otimes\omega^E_{m,g}}_1\nonumber\\
  \le&\N{\bar\omega^{RE}_{m,g}-\tau^R_m\otimes\bar\omega_{m,g}^E}_1&+2\N{\omega^{RE}_{m,g}-\bar\omega^{RE}_{m,g}}_1,\nonumber
\end{eqnarray}
which, in turns, implies that
\begin{equation}\nonumber
\begin{split}
\overline{\F}(\sS,m)
\ge 1&-\int\d g\N{\bar\omega^{RE}_{m,g}-\tau^R_m\otimes\bar\omega^E_{m,g}}_1\\
&-2\int\d g\N{\omega^{RE}_{m,g}-\bar\omega^{RE}_{m,g}}_1,
\end{split}
\end{equation}
for any choice of $\bar\omega^{RE}$ in
$\B(\omega^{RE}_\sS;\delta)$. Now, thanks to Lemma~3.2 of
Ref.~\cite{dec-capacity} and eq.~\reff{fidelity}, we know that
\begin{equation}\nonumber
  \int\d g\N{\omega^{RE}_{m,g}-\bar\omega^{RE}_{m,g}}_1\le \N{\bar\omega^{RE}-\omega^{RE}_\sS}_1\le2\delta,
\end{equation}
which leads us to the estimate
\begin{equation}\nonumber
  \overline{\F}(\sS,m)\ge 1-4\delta-\int\d g\N{\bar\omega^{RE}_{m,g}-\tau^R_m\otimes\bar\omega^E_{m,g}}_1.
\end{equation}
We are hence left with estimating the last group average. 

In order to do so, we exploit a technique used by
Renner~\cite{thesis} and Berta~\cite{mario}: by applying
Lemma~\ref{lemma:tr-hs}, for any given state $\sigma^E$ invertible on
$\supp\ \bar\omega^E$, we obtain the estimate
\begin{equation}\nonumber
\begin{split}
  \N{\bar\omega^{RE}_{m,g}-\tau^R_m\otimes\bar\omega^E_{m,g}}_1^2&\le m\Tr\left[(\bar\omega^{RE}_{m,g}-\tau^R_m\otimes\bar\omega_{m,g}^E)\ A_{m,g}^{RE}\right]\\
  &:=m\N{\tilde{\rho}^{RE}_{m,g}-\tau^R_m\otimes\tilde{\rho}^E_{m,g}}^2_2,
\end{split}
\end{equation}
where $A_{m,g}^{RE}:= (P^R_m\otimes\sigma^E)^{-1/2}(\bar\omega^{RE}_{m,g}-\tau^R_m\otimes\bar\omega_{m,g}^E)(P^R_m\otimes\sigma^E)^{-1/2}$,
$\N{X}_2:=\sqrt{\Tr[X^\dag X]}$ denotes the Hilbert-Schmidt
norm, and
\begin{equation}\nonumber
\tilde\rho^{RE}_{m,g}:=(P^R_m\otimes\sigma^E)^{-1/4}\bar\omega^{RE}_{m,g}(P^R_m\otimes\sigma^E)^{-1/4},
\end{equation}
and, correspondingly,
$\tilde\rho^E_{m,g}:=\Tr_R[\tilde\rho^{RE}_{m,g}]=(\sigma^E)^{-1/4}\bar\omega^E_{m,g}
(\sigma^E)^{-1/4}$. It is easy to check that
\begin{equation}\nonumber
  \N{\tilde\rho^{RE}_{m,g}-\tau^R_m\otimes\tilde\rho^E_{m,g}}^2_2=\N{\tilde\rho^{RE}_{m,g}}_2^2-\frac 1m\N{\tilde\rho^E_{m,g}}_2^2.
\end{equation}
Further, using the concavity of the function $f(x) = \sqrt{x}$, we have
\begin{equation}\label{eq:average1}
\begin{split}
&\phantom{\ge}\overline{\F}(\sS,m)\\
&\ge 1-4\delta -\sqrt{
\left\{m\int\d g\ \N{\tilde\rho^{RE}_{m,g}}_2^2-\int\d g\ \N{\tilde\rho^E_{m,g}}_2^2
\right\} 
}.
\end{split}
\end{equation}

Standard calculations, similar to those reported
in~\cite{state-merg,dec-capacity,mario}, lead to
\begin{equation}\nonumber
\int\d g\ \N{\tilde\rho^{RE}_{m,g}}_2^2=\frac sm\frac{s-m}{s^2-1}\N{\tilde\rho^{E}}_2^2+\frac sm\frac{ms-1}{s^2-1}\N{\tilde\rho^{RE}}_2^2
\end{equation}
and
\begin{equation}\nonumber
\int\d g\ \N{\tilde\rho^{E}_{m,g}}_2^2=\frac sm\frac{ms-1}{s^2-1}\N{\tilde\rho^E}_2^2+\frac sm\frac{s-m}{s^2-1}\N{\tilde\rho^{RE}}_2^2,
\end{equation}
where
\begin{equation}\nonumber
\tilde\rho^{RE}:=(\openone_R\otimes\sigma^E)^{-1/4}\bar\omega^{RE}(\openone_R\otimes\sigma^E)^{-1/4},
\end{equation}
and $\tilde\rho^E_\sS:=\Tr_R[\tilde\rho^{RE}_\sS]$. By simple
manipulations, we arrive at
\begin{equation}\nonumber
\begin{split}
  &\phantom{=}m\int\d g\ \N{\tilde\rho^{RE}_{m,g}}_2^2-\int\d g\ \N{\tilde\rho^E_{m,g}}_2^2\\
&=\frac{s^2(m^2-1)}{m(s^2-1)}\left\{\N{\tilde\rho^{RE}}_2^2-\frac 1s\N{\tilde\rho^E}_2^2\right\}.
\end{split}
\end{equation}
Since $m\le s$,
\begin{equation}\nonumber
  \frac{s^2(m^2-1)}{m(s^2-1)}=m\frac{1-\frac{1}{m^2}}{1-\frac{1}{s^2}}\le m,
\end{equation}
so that eq.~(\ref{eq:average1}) can be rewritten as
\begin{equation}\nonumber
  \overline{\F}(\sS,m)\ge 1-4\delta-\sqrt{
    m\left\{\N{\tilde\rho^{RE}}_2^2-\frac 1s\N{\tilde\rho^E}_2^2\right\}
  },
\end{equation}
for any choice of the states $\bar\omega^{RE}\in
\B(\omega^{RE}_\sS;\delta)$ and $\sigma^E$ invertible on $\supp\
\bar\omega^E$.

Now, notice that
\begin{equation}\nonumber
  \N{\tilde\rho^{RE}}^2_2\le2^{S_2(\bar\omega^{RE}\|\openone_R\otimes\sigma^E)}.
\end{equation}
This inequality easily follows from~\reff{orig}, i.e.,
\begin{equation}\nonumber
\begin{split}
  \Tr[(\omega^{-1/4}\rho\omega^{-1/4})^2]&=\Tr[\omega^{-1/2}\rho\omega^{-1/2}\rho]\\
  &\le\Tr[\rho^2\omega^{-1}]=2^{S_2(\rho\|\omega)}.
\end{split}
\end{equation}
Moreover, from Lemma~\ref{lemma:tr-hs}, $\N{\tilde\rho^E}_2^2\ge
1$. Thus,
\begin{equation}\nonumber
  \overline{\F}(\sS,m)\ge 1-4\delta-\sqrt{m\left\{2^{S_2(\bar\omega^{RE}\|\openone_R\otimes\sigma^E)}-\frac1s\right\}},
\end{equation}
for any choice of states $\bar\omega^{RE}\in
\B(\omega^{RE}_\sS;\delta)$ and $\sigma^E$, the latter strictly
positive on $\supp\ \bar\omega^{R}$. In order to tighten the bound, we
first optimize (i.e. minimize)
$S_2(\bar\omega^{RE}\|\openone_R\otimes\sigma^E)$ over $\sigma^E$ for
any $\bar\omega^{RE}$, obtaining $I^c_2(\bar\omega^{RE}|E)$. We
further optimize (i.e. minimize) $I^c_2(\bar\omega^{RE})$ over
$\bar\omega^{RE}\in\B(\omega^{RE}_\sS;\delta)$, eventually obtaining
$I^c_{2,\delta}(\omega^{RE}_\sS)$. $\blacksquare$\bigskip

\subsubsection{Proof of the lower bound in~\reff{eqcorr3}}

By Lemma~\ref{thm_one}, we have the following
\begin{corollary}
  Given a channel $\Phi:{\cal B}(\sH_A)\mapsto{\cal B}(\sH_B)$, an
  $s$-dimensional subspace $\sS\subseteq\sH_A$, and any
  $\delta\in[0,\eps/4]$, a non-negative real number $R=\log m$, $m\in\mathbb{N}$, is an
  $\eps$-achievable rate for entanglement transmission through $\Phi|_\sS$ if
\begin{equation}\nonumber
  4\delta+\sqrt{m\left\{2^{I^c_{2,\delta}(\omega^{RE}_\sS)}-\frac1s\right\}}\le\eps.\ \square
\end{equation}
\end{corollary}

In particular, since $s\le d:=\dim\sH_A$, a positive real number
$R=\log m$ is an $\eps$-achievable rate for $\Phi|_\sS$ if, for any
$\delta\in\left[0,\eps/4\right]$,
\begin{equation}\nonumber
  m2^{I^c_{2,\delta}(\omega^{RE}_\sS)}\le\frac 1d+(\eps-4\delta)^2,
\end{equation}
or, equivalently, if
\begin{equation}\nonumber
  \log m\le\log\left[\frac 1d+(\eps-4\delta)^2\right]-I^c_{2,\delta}(\omega^{RE}_\sS).
\end{equation}
This, together with the definition~\reff{coh-smooth}, implies the
following lower bound to the one-shot capacity of entanglement
transmission through $\Phi|_\sS$, for any $\delta\in\left[0,\eps/4\right]$:
\begin{equation}\nonumber
Q_{\ent}(\Phi|_\sS;\eps)\ge\log\left[\frac 1d+(\eps-4\delta)^2\right]+H_2^\delta(\omega^{RE}_\sS|E)-\Delta,
\end{equation}
where $\Delta\le 1$ is a positive quantity included to make the right hand side of the above inequality equal to the logarithm of a positive integer (see the Remark after Theorem~\ref{thm_main}). This in turn implies the following lower bound to the one-shot capacity of entanglement
transmission through $\Phi$:
\begin{equation}\nonumber
Q_{\ent}(\Phi;\eps)\ge\log\left[\frac 1d+(\eps-4\delta)^2\right]+\max_{\sS\subseteq\sH_A}H_2^\delta(\omega^{RE}_\sS|E) -\Delta.\end{equation}
As a consequence of
Lemma~\ref{lemma:s2dmax}, we have
\begin{equation}\nonumber
\begin{split}
&\phantom{\ge}  Q_{\ent}(\Phi;\eps)\\
&\ge\log\left[\frac 1d+(\eps-4\delta)^2\right]+\max_{\sS\subseteq\sH_A}H_{\min}^\delta(\omega^{RE}_\sS|E) -\Delta\\
  &\ge\log\left[\frac
    1d+(\eps-4\delta)^2\right]+\max_{\sS\subseteq\sH_A}H_{\min}^\delta(\omega^{RE}_\sS|\omega_\sS^E) -\Delta,
\end{split}
\end{equation}
where
\begin{equation}\nonumber
H_{\min}^\delta(\omega^{RE}_\sS|\omega_\sS^E)
:=-\min_{\bar\omega^{RE}\in\B(\omega^{RE}_\sS;\delta)}D_{\max}(\bar\omega^{RE}\|\openone_R\otimes\bar\omega^E),
\end{equation}
for $\bar\omega^E=\Tr_R[\bar\omega^{RE}]$. In~\cite{mario}, it is
proved that $H_{\min}(\rho^{AB}|\rho^B)=-H_0(\rho^{AC}|C)$, if
$\rho^{AB}$ and $\rho^{AC}$ are both reduced states of the same
tripartite pure state. This fact, together with arguments analogous to
those used in~\cite{roger} to prove Lemma~3 there, leads to
the identity $H_{\min}^{\delta}(\omega^{RE}_\sS|\omega_\sS^E)=
-H_0^{\delta}(\omega^{RB}_\sS|B)$, implying, via
definition~\reff{coh-smooth}, the desired lower bound to the one-shot
capacity of entanglement transmission:
\begin{equation}\label{eventaully}
  Q_{\ent}(\Phi;\eps)\ge\log\left[\frac 1d+(\eps-4\delta)^2\right]+\max_{\sS\subseteq\sH_A}I^c_{0,\delta}(\omega^{RB}_\sS) -\Delta,
\end{equation}
for any $\delta\in\left[0,\eps/4\right]$, and, in particular, for
$\delta=\eps/8$.

\subsection{Proof of the upper bound in Theorem \ref{thm_main}}
\label{upper}

In this section we prove the upper bound
\begin{equation}\nonumber
  Q_{\ent}(\Phi;\eps)\le\max_{\sS\subseteq\sH_A}\I_{0,2\sqrt{\eps}}(\omega^{RB}_\sS),
\end{equation}
where $\I_{0,2\sqrt{\eps}}(\omega^{RB}_\sS)$ is defined in
eq.~\reff{new-I0}.

We start by proving the following monotonicity relation: 

\begin{lemma}[Quantum data-processing inequality]\label{lemma:final}
  For any bipartite state $\rho^{AB}$, any channel $\Phi:B\mapsto C$, and any
  $\delta\ge 0$, we have
\begin{equation}\nonumber
  \I_{0,2\sqrt{\delta}}(\rho^{AB})\ge\I_{0,\delta}((\id\otimes \Phi)(\rho^{AB})).\ \square
\end{equation}
\end{lemma}

\noindent{\bf Proof.} Let
$P\in\P((\id\otimes\Phi)(\rho^{AB});\delta)$ and $\bar\sigma^C$ be
the pair achieving $\H_0^\delta((\id\otimes\Phi)(\rho^{AB})|C)$,
that is,
\begin{equation}\nonumber
\begin{split}
&\phantom{=}  \H_0^\delta((\id\otimes\Phi)(\rho^{AB})|C)\\
&=\log\Tr[\sqrt{P}\Pi_{(\id\otimes\Phi)(\rho^{AB})}\sqrt{P}(\openone_A\otimes\bar\sigma^C)].
\end{split}
\end{equation}
Consider now the operator
\begin{equation}\nonumber
Q:=(\id_A\otimes\Phi^*)
(\sqrt{P}\Pi_{(\id\otimes\Phi)(\rho^{AB})}\sqrt{P}),
\end{equation}
where $\Phi^*:C\mapsto B$ denotes the identity-preserving adjoint map
associated with the trace-preserving map $\Phi:B\mapsto C$. It clearly
satisfies $0\le Q\le\openone$. Let us now put, for sake of clarity,
$\gamma^{AC}:=(\id\otimes\Phi)(\rho^{AB})$. Then,
\begin{equation}\nonumber
\begin{split}
&\phantom{=}  \Tr[Q\ \rho^{AB}]\\
&=\Tr\left[\left(\sqrt{P}\Pi_{\gamma^{AC}}\sqrt{P}\right)\ \gamma^{AC}\right]\\
  &=1+\Tr\left[\Pi_{\gamma^{AC}}\left(\sqrt{P}\gamma^{AC}\sqrt{P}-\gamma^{AC}\right)\right]\\
  &\ge
  1+\Tr\left[\left\{\sqrt{P}\gamma^{AC}\sqrt{P}<\gamma^{AC}\right\}\left(\sqrt{P}\gamma^{AC}\sqrt{P}-\gamma^{AC}\right)\right],
\end{split}
\end{equation}
where in the last line we used Lemma~\ref{bowen}. Due to Gentle
Measurement Lemma~\ref{gm}, we have that
\begin{equation}\nonumber
\N{\sqrt{P}\gamma^{AC}\sqrt{P}-\gamma^{AC}}_1 \le2\sqrt{\delta},
\end{equation}
which, together with the formula $\N{A-B}_1=\Tr[\{A\ge
B\}(A-B)]-\Tr[\{A<B\}(A-B)]$, implies
\begin{equation}\nonumber
\begin{split}
&\phantom{\ge}  \Tr\left[\left\{\sqrt{P}\gamma^{AC}\sqrt{P}<\gamma^{AC}\right\}\left(\sqrt{P}\gamma^{AC}\sqrt{P}-\gamma^{AC}\right)\right]\\
&\ge -2\sqrt{\delta}.
\end{split}
\end{equation}
This leads to the estimate
\begin{equation}\nonumber
  \Tr[Q\ \rho^{AB}]\ge 1-2\sqrt{\delta}.
\end{equation}
In other words, $Q\in\P(\rho^{AB};2\sqrt{\delta})$. Now, let
$\bar\sigma^B$ be the state achieving
$\max_{\sigma^B\in\states(\sH_B)}\log\Tr[\sqrt{Q}\Pi_{\rho^{AB}}\sqrt{Q}\
(\openone_A\otimes\sigma^B)]$. We then have the following chain of
inequalities:
\begin{align}
&\H_0^\delta((\id\otimes\Phi)(\rho^{AB})|C)\nonumber\\
=&\log\Tr[\sqrt{P}\Pi_{(\id\otimes\Phi)(\rho^{AB})}\sqrt{P}(\openone_A\otimes\bar\sigma^C)]\nonumber\\
\ge&\log\Tr[\sqrt{P}\Pi_{(\id\otimes\Phi)(\rho^{AB})}\sqrt{P}(\openone_A\otimes\Phi(\bar\sigma^B))]\nonumber\\
  =&\log\Tr[Q\ (\openone_A\otimes\bar\sigma^B)]\nonumber\\
  \ge&\log\Tr[\sqrt{Q}\Pi_{\rho^{AB}}\sqrt{Q}\ (\openone_A\otimes\bar\sigma^B)]\nonumber\\
  =&\max_{\sigma^B\in\states(\sH_B)}\log\Tr[\sqrt{Q}\Pi_{\rho^{AB}}\sqrt{Q}\ (\openone_A\otimes\sigma^B)]\nonumber\\
  \ge&\H_0^{2\sqrt{\delta}}(\rho^{AB}|B) \nonumber.
\end{align}
The statement of the Lemma is finally obtained by
definition~\reff{new-I0}. $\blacksquare$\bigskip

\noindent With Lemma~\ref{lemma:final} in hand, it is now easy, by the
following standard arguments, to prove the upper bound in
Theorem~\ref{thm_main}.

In fact, suppose now that $R_0$ is the maximum of all
$\eps$-achievable rates, i.e., $R_0=Q_{\ent}(\Phi;\eps)$. By
Definition~\ref{eachrate}, the integer $s:=2^{R_0}$ is such that
\begin{equation}\nonumber
\F_{\ent}(\Phi;s)\ge 1-\eps.
\end{equation}
This is equivalent to saying that there exists an $s$-dimensional subspace
$\sS\subseteq\sH_A$ such that
\begin{equation}\nonumber
  \max_{\mD}F^2((\id_R\otimes\mD_B)(\omega^{RB}_\sS),\Psi^{RA}_\sS)\ge 1-\eps
\end{equation}
or, equivalently, that there exists a decoding operation
$\bar\mD:\mathcal{B}(\sH_B)\mapsto\mathcal{B}(\sH_A)$ such that
$\Psi^{RA}_{\sS}:=|\Psi^{RA}_{\sS}\>\<\Psi^{RA}_{\sS}|\in\P\left((\id_R\otimes\bar\mD_B)(\omega^{RB}_\sS);\eps\right)$.
Then, by exploiting Lemma~\ref{lemma:final}, we have that
\begin{equation}\nonumber
\begin{split}
&\I_{0,2\sqrt{\eps}}(\omega^{RB}_\sS)\\
\ge&\I_{0,\eps}((\id_R\otimes\bar\mD_B)(\omega^{RB}_\sS)\\ \ge&-\max_{\sigma^A}\log\Tr\left[\Psi^{RA}_{\sS}\Pi_{(\id_R\otimes\bar\mD_B)(\omega^{RB}_\sS)}\Psi^{RA}_{\sS}(\openone_R\otimes\sigma^A)\right]\\
  \ge&-\max_{\sigma^A}\log\Tr\left[\Psi^{RA}_{\sS}(\openone_R\otimes\sigma^A)\right].
\end{split}
\end{equation}
The claim is finally proved by noticing that the last line in the equation above equals $I_0^c(\Psi^{RA}_{\sS})$, so that $\I_{0,2\sqrt{\eps}}(\omega^{RB}_\sS)\ge I_0^c(\Psi^{RA}_{\sS}) =\log s=R_0=Q_{\ent}(\Phi;\eps)$.

\section{Quantum capacity of a sequence of channels}\label{multiple}

Let $\{\sH_A^{\otimes n}\}_{n=1}^\infty$ and $\{\sH_B^{\otimes
  n}\}_{n=1}^\infty$ be two sequences of Hilbert spaces, and let
$\hat\Phi:=\{\Phi_n\}_{n=1}^\infty$ be a sequence of quantum channels
such that, for each $n$,
\begin{equation}\nonumber
\Phi_n:\mathcal{B}(\sH_A^{\otimes n})\mapsto\mathcal{B}(\sH_B^{\otimes n}).
\end{equation} 
For any given $\eps>0$ and any fixed finite $n$, the one-shot quantum
capacity of $\Phi_n$, with respect to the fidelity $\F_x$, where
$x\in\{\ent,\min\}$, is given by $Q_x(\Phi_n;\eps)$. However,
since $\Phi_n$ itself could be the CPTP-map describing $n$ uses of an
arbitrary channel, possibly with memory, it is meaningful to introduce
the quantity
\begin{equation}\nonumber
  \frac 1nQ_x(\Phi_n;\eps),
\end{equation}
which can be interpreted as the capacity \emph{per use} of the
channel. This quantity is of relevance in all practical situations
because, instead of considering an asymptotically large number of uses
of the channel, it is more realistic to consider using a channel a
large \emph{but finite} number of times, in order to achieve reliable
transmission of quantum information. Theorem~\ref{thm_main} provides
the following bounds on this quantity:
\begin{equation}
\begin{split}\nonumber
\frac 1n&\max_{\sS\subseteq\sH_A^{\otimes n}}I^c_{0,\eps/8}(\omega^{R_nB_n}_\sS)+  \frac 1n\log\left[\frac
    1{d^n}+\frac{\eps^2}{4}\right]-\frac\Delta n\\
\le\frac 1n&\ Q_{\ent}(\Phi_n;\eps)\\
\le\frac 1n&\max_{\sS\subseteq\sH_A^{\otimes n}}\I_{0,2\sqrt{\eps}}(\omega^{R_nB_n}_\sS),
\end{split}
\end{equation}
where $\omega_{\sS_n}^{R_nB_n}=\Tr_{E_n}[\Omega^{R_nB_nE_n}_{\sS_n}]$,
the pure state $|\Omega^{R_nB_nE_n}_{\sS_n}\>$ being defined through
eq.~\reff{o-rbe}. Note that the second and third terms in the lower bound
decrease rapidly as $n$ increases, resulting in sharp bounds on the
capacity for entanglement transmission per use, even for finite
$n$. Moreover, due to Corollary~\ref{corr1}, the difference between
$Q_{\ent}(\Phi_n;\eps)/n$ and $Q_{\min}(\Phi_n;\eps)/n$ also decreases
as $n$ increases.

If the sequence is infinite, we define the corresponding
asymptotic capacity of the channel $\Phi$ as
\begin{equation}\nonumber
  Q_x^\infty(\hat\Phi):=\lim_{\eps\to 0}\liminf_{n\to\infty}\frac 1nQ_x(\Phi_n;\eps).
\end{equation}
Due to the equivalence relations stated in Corollary~\ref{corr1}, we
see that the different fidelities yield the same asymptotic quantum
capacity, so that
\begin{equation}\label{eq:qentqmin}
  Q_{\ent}^\infty(\hat\Phi)=Q_{\min}^\infty(\hat\Phi):=Q^\infty(\hat\Phi).
\end{equation}

\subsection{Multiple uses of a memoryless channel}\label{memless}

Here, we prove that the asymptotic quantum capacity of a memoryless
channel, sometimes referred to as the ``LSD Theorem''~\cite{lloyd,shor,devetak}, can be obtained from Theorem~\ref{thm_main}. For a memoryless channel, the sequence $\hat\Phi$ is given by $\{\Phi^{\otimes n}\}_{n=1}^\infty$, and hence its capacity can simply be labelled by $\Phi$.
The LSD Theorem, strictly speaking, gives an expression for $Q_{\min}^{\infty}(\Phi)$, whereas our method gives an expression for $Q_{\ent}^\infty(\Phi)$. However, by eq.~(\ref{eq:qentqmin}), these expressions are equivalent.

Here we prove the following theorem, which can be seen as an alternative formulation of the LSD theorem:
\begin{theorem}[memoryless channels]
\label{mless}
For a memoryless channel
$\Phi:\mathcal{B}(\sH_A)\mapsto\mathcal{B}(\sH_B)$,
\begin{equation}\label{eq:converse}
  Q^\infty(\Phi)=\lim_{n\to\infty}\frac 1n\max_{\sS\subseteq\sH_A^{\otimes n}}I^c(\sS,\Phi^{\otimes n}),
\end{equation}
where $I^c(\sS,\Phi)$ denotes the coherent information of the channel
$\Phi$ with respect to an input subspace $\sS$, and is defined
through~\reff{coh} as follows:
\begin{equation}\nonumber
I^c(\sS,\Phi):=I^c(\omega_\sS^{RB}),
\end{equation}
where $\omega_\sS^{RB}$ is the reduced state of the pure state
$|\Omega^{RBE}_\sS\>$ defined in~\reff{o-rbe}.  $\square$
\end{theorem}
Notice that in \reff{eq:converse} liminf has been replaced by lim, since the limit exists \cite{bar-niel-schum}.

\subsubsection{Direct part of Theorem~\ref{mless}}
Here we prove that 
\be  Q^\infty(\Phi)\ge \lim_{n\to\infty}\frac 1n\max_{\sS\subseteq\sH_A^{\otimes n}}I^c(\sS,\Phi^{\otimes n}),
\nonumber
\ee

From Theorem~\ref{thm_main}
\begin{equation}\nonumber
\begin{split}
 Q^\infty(\Phi)\ge\lim_{\eps\to 0}\lim_{n\to\infty}\frac 1n&\left\{
    \log\left[\frac 1{d^n}+\frac{\eps^2}{4}\right]-\Delta\right.\\
&\left.+\max_{\sS\subseteq\sH_A^{\otimes n}}I^c_{0,\eps/8}(\omega^{R_nB_n}_\sS)
  \right\}.
\end{split}
\end{equation}
The first two terms clearly vanish. We are hence left with the evaluation
of the third term. First of all, we recall that [see arguments before
eq.~\reff{eventaully}]
\begin{equation}\nonumber
  I^c_{0,\eps/8}(\omega^{R_nB_n}_\sS)= H_{\min}^{\eps/8}(\omega^{R_nE_n}_\sS|\omega^{E_n}_\sS).
\end{equation}
This implies that
\begin{equation}\nonumber
\begin{split}
Q^\infty(\Phi)&\ge\lim_{\eps\to 0}\lim_{n\to\infty}\frac 1n\max_{\sS\subseteq\sH_A^{\otimes n}}H_{\min}^{\eps/8}(\omega^{R_nE_n}_\sS|\omega^{E_n}_\sS)\\
&\ge\lim_{\eps\to 0}\lim_{n\to\infty}\frac 1n\max_{\sS\subseteq\sH_A}H_{\min}^{\eps/8}((\omega^{RE}_\sS)^{\otimes n}|(\omega^E_\sS)^{\otimes n}).\\
\end{split}
\end{equation}
As shown in~\cite{thesis}, we have
\begin{equation}\nonumber
\begin{split}
  &\phantom{=}\lim_{\eps\to 0}\lim_{n\to\infty}\frac 1n\max_{\sS\subseteq\sH_A}H_{\min}^{\eps/8}((\omega^{RE}_\sS)^{\otimes n}|(\omega^E_\sS)^{\otimes n})\\
&=\max_{\sS\subseteq\sH_A}H(\omega^{RE}_\sS|\omega^E_\sS)\\
  :&=\max_{\sS\subseteq\sH_A}[-I^c(\omega^{RE}_\sS)]\\
  &=\max_{\sS\subseteq\sH_A}I^c(\sS,\Phi),
\end{split}
\end{equation}
where in the last line we used the fact that $I^c(\omega_\sS^{RB})=-I^c(\omega_\sS^{RE})$, since
$\Omega^{RBE}_\sS$ is pure. Therefore,
\begin{equation}\nonumber
  Q^\infty(\Phi)\ge\max_{\sS\subseteq\sH_A}I^c(\sS,\Phi).
\end{equation}
As in~\cite{bar-niel-schum}, we can then achieve the right hand side
of~\reff{eq:converse} by the usual blocking argument.

\subsubsection{Weak converse of Theorem~\ref{mless}}

In order to obtain the upper bound, it suffices to evaluate the
asymptotic behaviour of the upper bound on $Q^\infty(\Phi)$
which, by Theorem~\ref{thm_main}, is given by
\begin{equation}\label{ubp}
 Q^\infty(\Phi)\le\lim_{\eps\to 0}\lim_{n \rightarrow \infty}\frac 1n\max_{\sS\subseteq\sH_A^{\otimes n}}\I_{0,2\sqrt{\eps}}(\omega_\sS^{R_nB_n})).
\end{equation}

The following two lemmas are essential for the evaluation of this bound, and are also
of independent interest. The first one relates $ S^P_1(\rho\|\sigma)$ to the
quantum relative entropy $S(\rho\|\sigma)$, while the second one relates the 
operator-smoothed $0$-coherent information to the usual coherent information.
\begin{lemma}\label{lem:S1P}
Consider two states $\rho, \sigma \in \states({\cal{H}})$, with
$\supp\rho\subseteq\supp\sigma$, and a positive operator $0\le
P\le\openone$ such that $P\in\P(\rho;\delta)$ for some given 
$\delta \ge 0$. Then we have
\begin{equation}\label{eqlems1p}
  S^P_1(\rho\|\sigma)\le\frac{S(\sqrt{P}\rho\sqrt{P}\|\sigma)+2\delta^\prime\log d+2}{1-\delta^\prime},
\end{equation}
where $\delta^\prime := 2\sqrt{\delta}$, and $d:= \dim {\cal{H}}$. $\square$ 
\end{lemma}
{\bf Proof.}  The main ingredients of the proof of this lemma are the
monotonicity property of the operator-smoothed conditional entropy
(Lemma~\ref{lemma:h-mono}), the matrix convexity of the function
$t\log t$, and the Fannes' inequality. From~\reff{S1P}, we have:
\begin{eqnarray}
 & & S^P_1(\rho\|\sigma)\nonumber\\
&=&\frac{\Tr\left[\sqrt{P}\rho\log\rho\sqrt{P}\ \Pi_\sigma-\sqrt{P}\rho\sqrt{P}\log\sigma\right]}{\Tr[\sqrt{P}\rho\sqrt{P}\ 
\Pi_\sigma]}\nonumber\\
&=& \frac{\Tr[\rho\log\rho]-\Tr[(\openone-P')(\rho\log\rho)]-\Tr[\sqrt{P}\rho\sqrt{P}\log\sigma]}{\Tr[P'\rho]},\nonumber\\
\end{eqnarray}
where, for our convenience, we have put
$P':=\sqrt{P}\Pi_\sigma\sqrt{P}$. Since $P\in\P(\rho;\delta)$, due to
Lemma~\ref{gm},
\begin{equation}\label{asdfgh}
\N{\rho-\sqrt{P}\rho\sqrt{P}}_1\le \delta^\prime,
\end{equation}
where $\delta^\prime:=2\sqrt{\delta}$. Obviously $0\le P' \le \openone$.
Using \reff{asdfgh}, the fact that $\openone \ge \Pi_\sigma\ge\Pi_\rho$, and the cyclicity
of the trace, we have
\begin{eqnarray}
\tr[P'\rho] &=& \tr [\sqrt{P}\Pi_\sigma \sqrt{P}\ \rho]\nonumber\\
&=& \tr [\Pi_\rho \sqrt{P}\rho\sqrt{P}]\nonumber\\
&=& \tr [\Pi_\rho\ \rho] -\tr \left[\Pi_\rho(\rho -  \sqrt{P}\rho\sqrt{P})\right]\nonumber\\
&\ge & 1 - \delta'.
\end{eqnarray}
Hence, $P'\in\P(\rho;\delta^\prime)$.

Since $t\log t$ is a matrix convex function, it is known that
\begin{equation}\nonumber
-\Tr[K^\dag(\rho\log\rho)K]\le-\Tr[(K^\dag\rho K)\log(K^\dag\rho K)],
\end{equation}
for any contraction $K$,~\cite{bhatia}. Let
$K=K^\dag=\sqrt{\openone-P'}$. Then
\begin{equation}\nonumber
\Tr[(\openone-P')(-\rho\log\rho)]\le S(\bar\rho),
\end{equation}
where $\bar\rho$ is the sub-normalized density matrix defined as
$\bar\rho:=\sqrt{\openone-P'}\rho\sqrt{\openone-P'}$. It is clear that,
since $P'\in\P(\rho;\delta^\prime)$,
$\Tr[\bar\rho]\le\delta^\prime$. Moreover, by simple algebra,
$S(\bar\rho)\le\delta^\prime\log d+1$. This implies that
\begin{equation}
\begin{split}
  &\phantom{\le}  S^P_1(\rho\|\sigma)\\
  &\le\frac{\Tr[\rho\log\rho]-\Tr[\sqrt{P}\rho\sqrt{P}\log\sigma]+\delta'\log
    d+1}{1-\delta'}.
\end{split}
\end{equation}

By~\reff{asdfgh} and Fannes' continuity property of the von Neumann
entropy~\cite{haya}, we have that
\begin{equation}\nonumber
\begin{split}
  &\phantom{\le}  \Tr[\rho\log\rho]\\
  &\le\Tr\left[\sqrt{P}\rho\sqrt{P}\log(\sqrt{P}\rho\sqrt{P})\right]+\delta'\log
  d+1,
\end{split}
\end{equation}
which in turn yields \reff{eqlems1p}.
$\blacksquare$

\begin{lemma}\label{lem:cohinf}
For any bipartite state $\rho^{AB}\in \states({\cal{H}}_A \otimes {\cal{H}}_B)$, and any given $\delta \ge 0$, we have
\begin{equation}\label{statement}
  \I_{0,\delta}(\rho^{AB})\le\frac{ I^c(\rho^{AB})}{1-\delta^\prime}
+\frac{4(\delta^\prime\log (d_Ad_B)+1)}{1-\delta^\prime}, 
\end{equation}
where $d_A := \dim {\cal{H}}_A$, $d_B := \dim {\cal{H}}_B$, and $\delta^\prime := 2\sqrt{\delta}$. $\square$
\end{lemma}
{\bf{Proof.}}
By Lemma \ref{lemma:h-mono} we have 
\begin{eqnarray}\label{step1}
  \I_{0,\delta}(\rho^{AB})&\le&\I_{1,\delta}(\rho^{AB})\nonumber\\
&=&\max_{P\in\P(\rho^{AB};\delta)}\min_{\sigma^{B}}S^P_1(\rho^{AB}\|\tau^{A}\otimes\sigma^{B})\nonumber\\
 & & \quad \quad - \log d_A,
\end{eqnarray}
where $\tau^A := \openone_A/d_A$, namely, the completely mixed state. In the
above, we have made use of the following identity, which is easily obtained from \reff{S1P}: 
for two states $\rho$ and $\sigma$, and any constant
$c>0$, $S^P_1(\rho\|c\sigma)=S^P_1(\rho\|\sigma)-\log c$.
Using Lemma \ref{lem:S1P}, and the analogous identity, $S(\rho\|c\sigma)=S(\rho\|\sigma)-\log c$, we have: for $\delta^\prime = 2 
\sqrt{\delta}$,
\begin{eqnarray}\label{step2}
& & S^P_1(\rho^{AB}\|\tau^{A}\otimes\sigma^{B})\nonumber\\
&\le & \frac{S(\sqrt{P}\rho^{AB}\sqrt{P}\|\tau^A \otimes \sigma^B)+2\delta^\prime\log (d_Ad_B)+2}{1-\delta^\prime}\nonumber\\
&=& \frac{S(\sqrt{P}\rho^{AB}\sqrt{P}\|\openone^A \otimes \sigma^B)+2\delta^\prime\log (d_Ad_B)+2}{1-\delta^\prime} 
\nonumber\\
& & \quad \quad \quad + \frac{1}{1-\delta^\prime} \log d_A.\nonumber\\
\end{eqnarray}
Then by \reff{step1}, \reff{step2}, and Lemma \ref{lemma:qmi_as_min}, we obtain 
\begin{eqnarray}\label{step3}
  \I_{0,\delta}(\rho^{AB})&\le&\frac{ I^c(\sqrt{P}\rho^{AB}\sqrt{P})}{1-\delta^\prime}
+\frac{2\delta^\prime\log (d_Ad_B)+2}{1-\delta^\prime} 
\nonumber\\
& & \quad \quad \quad + \frac{\delta^\prime}{1-\delta^\prime} \log d_A.\nonumber\\
\end{eqnarray}
Finally, applying Fannes' inequality to each of the terms on the right hand side of the
identity $I^c({{\omega^{AB}}})= S({{\omega^{B}}})- S({{\omega^{AB}}})$, 
where ${{\omega^{AB}}}:=\sqrt{P}{{\rho^{AB}}}\sqrt{P}$, and 
${{\omega^{B}}}:= \tr_A{{\omega^{AB}}}$, we obtain \reff{statement}.
$\blacksquare$
\medskip

From \reff{ubp} and  Lemma~\ref{lem:cohinf} we obtain 
\begin{equation}
\begin{split}
Q^\infty(\Phi) &\le \lim_{n\to\infty}\frac 1n\max_{\sS\subseteq\sH_A^{\otimes n}}I^c(\omega_\sS^{R_nB_n})\nonumber\\
&= \lim_{n\to\infty}\frac 1n\max_{\sS\subseteq\sH_A^{\otimes
    n}}I^c(\sS,\Phi^{\otimes n}),
\end{split}
\end{equation}
as claimed.

\subsection{Multiple uses of an arbitrary channel}\label{memwith}

To evaluate the quantum capacity of an arbitrary sequence of channels,
we employ the well-known Quantum Information Spectrum
Method~\cite{info-spect,hayashi-naga}. Two fundamental quantities used
in this approach are the \emph{quantum spectral sup}- and
\emph{inf-divergence rates}, defined as follows:
\begin{definition}[Spectral Divergence Rates]
  Given a sequence of states $\hat\rho=\{\rho_n\}_{n=1}^\infty$ and a
  sequence of positive operators
  $\hat\sigma=\{\sigma_n\}_{n=1}^\infty$, the quantum spectral
  sup- (inf-)divergence rates are defined in terms of the difference
  operators $\Pi_n(\gamma) = \rho_n - 2^{n\gamma}\sigma_n$ as
\begin{align}
  \overline{D}(\hat\rho \| \hat\sigma) &:= \inf \left\{ \gamma : \limsup_{n\rightarrow \infty} \mathrm{Tr}\left[ \{ \Pi_n(\gamma) \geq 0 \} \Pi_n(\gamma) \right] = 0 \right\} \label{odiv} \\
  \underline{D}(\hat\rho \| \hat\sigma) &:= \sup \left\{ \gamma :
  \liminf_{n\rightarrow \infty} \mathrm{Tr}\left[ \{ \Pi_n(\gamma) \geq
  0 \} \Pi_n(\gamma) \right] = 1 \right\} \label{udiv}
\end{align}
respectively.  $\square$
\end{definition}

It is known that (see e.~g.~\cite{bowen-datta})
\begin{equation}\label{rell}
  \overline{D}(\hat\rho \| \hat\sigma) \ge\underline{D}(\hat\rho \| \hat\sigma). 
\end{equation}

In analogy with the usual definition of the coherent information~\reff{coh}, we moreover define the \emph{spectral sup-} and \emph{inf-coherent information rates}, respectively, as follows: \begin{align}
  \overline{I}^c(\hat\rho^{RB})&:=\min_{\hat\sigma^B}\overline{D}(\hat\rho^{RB}\|\hat\openone_R\otimes\hat\sigma^B),\label{strong-conv}\\
\underline{I}^c(\hat\rho^{RB})&:=\min_{\hat\sigma^B}\underline{D}(\hat\rho^{RB}\|\hat\openone_R\otimes\hat\sigma^B),\label{att}
\end{align}
where $\hat\rho^{RB}:=\{\rho^{R_nB_n}\in\states(\sH_R^{\otimes
  n}\otimes\sH_B^{\otimes n})\}_{n=1}^\infty$,
$\hat\sigma^B:=\{\sigma^{B_n}\in\states(\sH_B^{\otimes
  n})\}_{n=1}^\infty$, and $\hat\openone_R:=\{
\openone_{R_n}\}_{n=1}^\infty$. The inequality~\reff{rell} ensures that
\begin{equation}
  \overline{I}^c(\hat\rho^{RB})\ge\underline{I}^c(\hat\rho^{RB}).
\end{equation}
Note that in eq.~\reff{strong-conv} and~\reff{att} we could write
minimum instead of infimum due to Lemma~1 in~\cite{hayashi-naga}. The
same remark applies also to the following:

\begin{theorem}[arbitrary channels]\label{thm3}
  The quantum capacity of $\hat\Phi$ is given by
\begin{equation}\nonumber
  Q^\infty(\hat\Phi)=\max_{\hat\sS}\underline{I}^c(\hat\omega_{\hat\sS}^{RB}),
\end{equation}
where $\hat\sS:=\{\sS_n:\sS_n\subseteq\sH_A^{\otimes
  n}\}_{n=1}^\infty$, and
$\hat\omega_{\hat\sS}^{RB}:=\{\omega_{\sS_n}^{R_nB_n}\}_{n=1}^\infty$,
with $\omega_{\sS_n}^{R_nB_n}=\Tr_{E_n}[\Omega^{R_nB_nE_n}_{\sS_n}]$,
the pure state $|\Omega^{R_nB_nE_n}_{\sS_n}\>$ being defined through
eq.~\reff{o-rbe}.  $\square$
\end{theorem}

The above theorem follows directly from Theorem~\ref{thm_main} and
Lemma~\ref{lemma:x} and Lemma~\ref{lemma:y} given below.

\begin{lemma}[Direct part]\label{lemma:x}
  Given a sequence of bipartite states $\hat\rho^{RB}$,
\begin{equation}\nonumber
\begin{split}
  \lim_{\delta\to 0}\liminf_{n\to\infty}&\max_{\bar\rho^{R_nB_n}_n\in\B(\rho^{R_nB_n}_n;\delta)}\min_{\sigma^{B_n}_n}\frac 1nS_0(\bar\rho_n^{R_nB_n}\|\openone_{R_n}\otimes\sigma^{B_n}_n)\\
&\ge\min_{\hat\sigma^B}\underline{D}(\hat\rho^{RB}\|\hat\openone_R\otimes\hat\sigma^B).\ \square
\end{split}
\end{equation}
\end{lemma}

\noindent{\bf Proof.} This follows directly from Theorem~3
of~\cite{nila}. $\blacksquare$\bigskip

\begin{lemma}[Weak converse]\label{lemma:y}
  Given a sequence of bipartite states $\hat\rho^{RB}$,
\begin{equation}\nonumber
\begin{split}
  \lim_{\delta\to 0}\liminf_{n\to\infty}&\max_{P_n\in\P(\rho^{R_nB_n}_n;\delta)}\min_{\sigma^{B_n}_n}\frac 1nS_0^{P_n}(\rho_n^{R_nB_n}\|\openone_{R_n}\otimes\sigma^{B_n}_n)\\
&\le\min_{\hat\sigma^B}\underline{D}(\hat\rho^{RB}\|\hat\openone_R\otimes\hat\sigma^B).\ \square
\end{split}
\end{equation}
\end{lemma}

\noindent{\bf Proof.} The proof is by \emph{reductio ad absurdum}: we
will assume that
\begin{equation}\label{ass}
\begin{split}
  \lim_{\delta\to 0}\liminf_{n\to\infty}&\max_{P_n\in\P(\rho^{R_nB_n}_n;\delta)}\min_{\sigma^{B_n}_n}\frac 1n S_0^{P_n}(\rho_n^{R_nB_n}\|\openone_{R_n}\otimes\sigma^{B_n}_n)\\
&>\min_{\hat\sigma^B}\underline{D}(\hat\rho^{RB}\|\hat\openone_R\otimes\hat\sigma^B),
\end{split}
\end{equation}
and show that such an assumption leads to a contradiction, hence
proving the statement of the lemma.

Let $\hat{\bar\sigma}:=\{\bar\sigma_n^{B_n}\}_{n=1}^\infty$ be the
sequence achieving $\min_{\hat\sigma^B}\underline{D}
(\hat\rho^{RB}\|\hat\openone_R\otimes\hat\sigma^B)$. Moreover, for any
$\delta>0$ fixed but arbitrary, let $\{\bar P_n\}_{n=1}^\infty$ be the
sequence of operators, satisfying both $0\le \bar
P_n\le\openone_{R_nB_n}$ and $\Tr[\bar
P_n\rho_n^{R_nB_n}]\ge1-\delta$, achieving the maximum over $P_n$ of
$\min_{\sigma^{B_n}_n} S_0^{P_n}(\rho_n^{R_nB_n}\|
\openone_{R_n}\otimes\sigma^{B_n}_n)$, for all $n$. Then,
eq.~\reff{ass} implies that
\begin{equation}\label{ass2}
\lim_{\delta\to 0}\liminf_{n\to\infty}\frac 1n S_0^{\bar P_n}(\rho_n^{R_nB_n}\|\openone_{R_n}\otimes\bar\sigma^{B_n}_n)>\underline{D}(\hat\rho^{RB}\|\hat\openone_R\otimes\hat{\bar\sigma}^B).
\end{equation}
 By arguments analogous to those used in the proof of
Lemma~\ref{lemma:final}, we can see that
\begin{equation}\label{contra}
  \tr\left[\sqrt{\bar P_n}\Pi_{\rho_n^{R_nB_n}}\sqrt{\bar P_n}\ \rho_n^{R_nB_n}\right]\ge 1-2\sqrt{\delta}.
\end{equation}
For our convenience, let us put
\begin{equation}\nonumber
\begin{split}
\beta_\delta&:=\liminf_{n\to\infty}\frac 1nS_0^{\bar P_n}(\rho_n^{R_nB_n}\|\openone_{R_n}\otimes\bar\sigma^{B_n}_n),\\
\gamma&:=\lim_{\delta\to0}\beta_\delta.
\end{split}
\end{equation}
It is clear that $\beta_\delta\ge\gamma$. Now, the assumption~\reff{ass}
implies~\reff{ass2}, which is in turn equivalent to
\begin{equation}\nonumber
  \gamma>\min_{\hat\sigma^B}\underline{D}(\hat\rho^{RB}\|\hat\openone_R\otimes\hat\sigma^B).
\end{equation}
Let then $\gamma_0$ be such that
\begin{equation}\nonumber
  \beta_\delta>\gamma_0>\min_{\hat\sigma^B}\underline{D}(\hat\rho^{RB}\|\hat\openone_R\otimes\hat\sigma^B).
\end{equation}
Moreover, by the definition of $\liminf$, there exists an $n_0$ such
that, for all $n\ge n_0$,
\begin{equation}\nonumber
  \frac 1nS_0^{\bar P_n}(\rho_n^{R_nB_n}\|\openone_{R_n}\otimes\bar\sigma^{B_n}_n)\ge\beta_\delta.
\end{equation}
The above equation can be rewritten as
\begin{equation}\nonumber
  \Tr\left[\sqrt{\bar P_n}\Pi_{\rho_n^{R_nB_n}}\sqrt{\bar P_n}\left(\openone_{R_n}\otimes\bar\sigma^{B_n}_n\right)\right]\le 2^{-n\beta_\delta},
\end{equation}
for all $n\ge n_0$. Now, for all $n\ge n_0$,
\begin{eqnarray}
  &&\Tr\left[\sqrt{\bar P_n}\Pi_{\rho_n^{R_nB_n}}\sqrt{\bar P_n}\
    \rho_n^{R_nB_n}\right]\nonumber\\
  &=&\Tr\bigg[\sqrt{\bar P_n}\Pi_{\rho_n^{R_nB_n}}\sqrt{\bar      P_n}\bigg(\rho_n^{R_nB_n}-2^{n\gamma_0}\left(\openone_{R_n}\otimes\bar\sigma^{B_n}_n\right)\bigg)\bigg]\nonumber\\
&&+2^{n\gamma_0}\Tr\left[\sqrt{\bar
      P_n}\Pi_{\rho_n^{R_nB_n}}\sqrt{\bar
      P_n}(\openone_{R_n}\otimes\bar\sigma^{B_n}_n)\right]\nonumber\\ &\le&\Tr\bigg[\{\rho_n^{R_nB_n}\ge2^{n\gamma_0}(\openone_{R_n}\otimes\bar\sigma^{B_n}_n)\}\nonumber\\
&&\phantom{+++++++}\times\bigg(\rho_n^{R_nB_n}-2^{n\gamma_0}\left(\openone_{R_n}\otimes\bar\sigma^{B_n}_n\right)\bigg)\bigg]\nonumber\\
&&+2^{-n(\beta_\delta-\gamma_0)},\nonumber
\end{eqnarray}
where, in the last step, we used Lemma~\ref{bowen}. The second term in
the sum goes to $0$ as $n\to\infty$, since we chose
$\gamma_0<\beta_\delta$. The first term, on the other hand, has to be
bounded away from $1$ due to the definition~\reff{udiv}, since
$\gamma_0>\underline{D}(\hat\rho^{RB}\|
\hat\openone_R\otimes\hat{\bar\sigma}^B)$. Hence the
assumption~\reff{ass} leads to
\begin{equation}\nonumber
\liminf_{n\to\infty}\Tr\left[\sqrt{\bar P_n}\Pi_{\rho_n^{R_nB_n}}\sqrt{\bar P_n}\
    \rho_n^{R_nB_n}\right]\le 1-c_0,
\end{equation}
where $c_0>0$ is a constant independent of $\delta$. This is clearly in
contradiction with~\reff{contra}, which holds for all $n$ and any
arbitrary $\delta>0$. $\blacksquare$

\section{Discussion}\label{discussion}

In this paper we obtained bounds on the one-shot entanglement transmission capacity of an arbitrary quantum channel, which itself could correspond to a finite number of uses of a channel with arbitrarily correlated noise. Our result, in turn, yielded bounds on the one-shot quantum capacity of the channel. Further, for multiple uses of a memoryless channel, our results led to an expression for the asymptotic quantum capacity of the channel, in terms of the regularized coherent information. This provided an alternative form of the LSD theorem, which was however known to be equivalent to it~\cite{dennis}. Finally, by employing the Quantum Information Spectrum Method, we obtained an expression for the quantum capacity of an arbitrary infinite sequence of channels.

\section*{Acknowledgments}

We would like to thank Mario Berta for providing us with a copy of his
Diploma thesis. We are also grateful to Fernando Brandao, Roger
Colbeck, Renato Renner, and Marco Tomamichel for interesting
discussions. The research leading to these results has received
funding from the European Community's Seventh Framework Programme
(FP7/2007-2013) under grant agreement number 213681.

\appendix

\section*{Biographies}
\medskip

\noindent
Francesco Buscemi received the Ph.D. in Physics from the University of Pavia, Italy, in 2006.

From 2006 to 2008, he was Researcher in quantum information theory at the Japan Science and Technology Agency, ERATO-SORST Quantum Computation and Information Project, Tokyo, Japan. At the end of 2008 he joined the University of Cambridge, U.K., as Research Associate at the Statistical Laboratory.
\medskip

\noindent
Nilanjana Datta received a Ph.D. degree from ETH Zurich,
Switzerland, in 1996. 

From 1997 to 2000, she was a 
postdoctoral researcher at the Dublin Institute of Advanced 
Studies, C.N.R.S. Marseille, and EPFL in Lausanne. In 2001
she joined the University of Cambridge, as a Lecturer in
Mathematics of Pembroke College, and a member of the Statistical
Laboratory, in the Centre for Mathematical Sciences. She is 
currently an Affiliated Lecturer of the Faculty of Mathematics,
University of Cambridge, and a Fellow of Pembroke College.
\medskip

\end{document}